\documentclass[prd,twocolumn,floatfix,showpacs,preprintnumbers,nofootinbib,superscriptaddress]{revtex4}
\usepackage[utf8x]{inputenc}
\usepackage{mathrsfs}
\usepackage{amssymb}
\usepackage{amsmath}
\usepackage{graphicx}

\usepackage[caption = false]{subfig}
\usepackage[normalem]{ulem}
\usepackage{xcolor}
\definecolor{lcolor}{rgb}{0.5,0,0}
\definecolor{citcolor}{rgb}{0,0.3,0.0}

\usepackage[breaklinks,colorlinks,urlcolor=blue,citecolor=citcolor,linkcolor=lcolor]{hyperref}
\usepackage{mciteplus}

\newcommand{\rt}{{\mathbf{r}}}
\newcommand{\xt}{{\mathbf{x}}}
\newcommand{\bt}{{\mathbf{b}}}
\newcommand{\bti}{{\mathbf{b}_{i}}}
\newcommand{\yt}{{\mathbf{y}}}
\newcommand{\zt}{{\mathbf{z}}}

\newcommand{\kt}{{\mathbf{k}}}

\newcommand{\Deltat}{{\boldsymbol{\Delta}}}

\newcommand{\tr}{\, \mathrm{Tr} \, }

\newcommand{\nc}{{N_\mathrm{c}}}
\newcommand{\nf}{{N_\mathrm{f}}}

\newcommand{\gev}{\ \textrm{GeV}}
\newcommand{\fm}{\ \textrm{fm}}

\newcommand{\lqcd}{\Lambda_{\mathrm{QCD}}}
\newcommand{\as}{\alpha_{\mathrm{s}}}
\newcommand{\aem}{\alpha_{\mathrm{em}}}

\newcommand{\xpom}{{x_\mathbb{P}}}

\newcommand{\der}{\mathrm{d}}

\newcommand{\A}{{\mathcal{A}}}

\begin{document}

\author{Heikki Mäntysaari}
\affiliation{
Department of Physics, P.O. Box 35, 40014 University of Jyv\"askyl\"a, Finland
}

\author{Bj\"orn Schenke}
\affiliation{
Physics Department, Brookhaven National Laboratory, Upton, NY 11973, USA
}

\title{
Confronting impact parameter dependent JIMWLK evolution with HERA data
}


\preprint{}

\begin{abstract}
The small-$x$ evolution of protons is determined from numerical solutions of the JIMWLK equations, starting from an initial condition at moderate $x$ for a finite size proton. The resulting dipole amplitude is used to calculate the total reduced cross section $\sigma_r$ and charm reduced cross section $\sigma_{rc}$, as well as diffractive vector meson production. We compare results to experimental data from HERA and discuss fundamental problems arising from the regime sensitive to non-perturbative physics. We emphasize that information on the gluonic content of the proton, gluon spatial distributions and correlations over wide ranges in $x$, which can in principle be constrained by our study, are essential ingredients for describing the initial state in proton-proton and proton-ion collisions. Future electron nucleus collisions at an electron-ion collider will provide important additional insight for heavier nuclei. We further demonstrate that it is not possible to probe gluon saturation in electron-proton collision at HERA energies and that electron-heavy ion collisions will be essential to access the saturation regime. \end{abstract}

\maketitle

\section{Introduction}

Deep inelastic scattering (DIS) is a clean and powerful process to explore the structure of hadrons as a function of longitudinal momentum fraction $x$ and distance scale $Q^2$. The most precise DIS measurements so far have been performed at the DESY-HERA electron-proton collider, and the HERA experiments H1 and ZEUS have recently published their  precise combined measurements of the proton structure functions~\cite{Aaron:2009aa,Abramowicz:1900rp,Abramowicz:2015mha,H1:2018flt}. These measurements make it possible to constrain the total parton densities inside the proton to a very good precision. 
In addition, measurements of diffractive vector meson production~\cite{Aktas:2005xu,Chekanov:2002xi,Chekanov:2002rm,Aktas:2003zi,Chekanov:2009ab,Breitweg:1999jy,Aaij:2013jxj,TheALICE:2014dwa,CMS:2016nct} provide information on the spatial extent of the gluon distribution and in case of incoherent diffraction, even its fluctuations \cite{Miettinen:1978jb, Mantysaari:2016ykx, Mantysaari:2016jaz}. In the future, if an electron-ion collider is realized~\cite{Boer:2011fh,Accardi:2012qut,Aschenauer:2017jsk}, DIS with a variety of nuclear targets will allow unprecedented detailed studies of proton and nuclear structure (see for example~\cite{Toll:2012mb,Lappi:2014foa,Hatta:2016dxp,Aschenauer:2017oxs}).

A successful framework for describing DIS at high energy is given by the dipole model \cite{Nikolaev:1990ja,Mueller:1993rr,Nikolaev:1994ce,Mueller:1994jq,Mueller:1994gb}. Here, the scattering process is studied in the rest frame of the target proton or nucleus, where the virtual photon first splits into a quark-antiquark pair (the color dipole), which subsequently interacts with the target.
Furthermore, at high enough energy we can assume the transverse distance between the quark and antiquark, $r$, to be constant during the interaction with the target.

The properties of the target at small enough longitudinal momentum fraction $x$ (high enough energy) enter via Wilson lines \cite{Balitsky:2001gj}, which can be computed in the color glass condensate framework \cite{Iancu:2003xm,Gelis:2010nm} and descibe the color rotation of the incoming quark or anti-quark. The interaction of the dipole with the target is thus described by a two-point function expressed by the Wilson lines at the position of the quark and anti-quark.\footnote{The interaction of the anti-quark is described by the Hermitian conjugate Wilson line.} This two-point function is the dipole operator, whose average over color configurations yields the dipole amplitude
\begin{equation}\label{eq:dipole_wilsonline_intro}
	N\left( \rt, \bt , x  \right) = \left \langle 1 - \frac{1}{\nc} \tr \left( V(\xt,x)V^\dagger(\yt,x) \right) \right \rangle.
\end{equation}
Here, $\bt=(\xt+\yt)/2$ is the impact parameter and $\rt=\xt-\yt$ represents the size and orientation of the color dipole.

There are several calculations in the color glass condensate framework on the market that either parametrize the dipole amplitude and its $x$ and geometry dependence or perform Balitsky-Kovchegov (BK) small $x$ evolution \cite{Balitsky:1995ub,Kovchegov:1999yj} of the dipole amplitude in the large $N_c$ limit, see e.g. \cite{Kowalski:2003hm,Iancu:2003ge,Kowalski:2006hc,Albacete:2010sy,Berger:2011ew,Rezaeian:2012ji,Rezaeian:2013tka,Lappi:2013zma,Albacete:2015xza,Iancu:2015joa}. 

There have however not been any calculations of DIS structure functions, or diffractive vector meson production on the level of Wilson lines including the full leading logarithmic JIMWLK evolution \cite{JalilianMarian:1996xn,JalilianMarian:1997jx, JalilianMarian:1997gr,Iancu:2001md, Ferreiro:2001qy, Iancu:2001ad, Iancu:2000hn}. The JIMWLK equation is a renormalization group equation for the Wilson lines, obtained by integrating out the quantum fluctuations at smaller and smaller Bjorken-$x$. Writing the JIMWLK equation in its stochastic Langevin form provides a convenient picture of the small-$x$ evolution as a random walk in color space and allows numerical solutions. The first numerical results were obtained for infinite nuclei in Refs.~\cite{Rummukainen:2003ns,Kovchegov:2008mk}, and the finite proton geometry was studied in \cite{Schlichting:2014ipa}.

In this work we perform for the first time calculations of proton structure functions and diffractive vector meson production using this most fundamental description of the target proton by its JIMWLK evolved Wilson lines.\footnote{In \cite{Kuokkanen:2011je} a Gaussian truncation of the JIMWLK equations was used (at NLO) to produce comparisons with HERA data. However, this is similar to the BK truncation and does not involve the exact evolution of the JIMWLK equations as performed in this work.} This has various advantages over simple parametrization models or those invoking BK evolution. For example, our description includes the expected growth of the proton with decreasing $x$ \cite{Schlichting:2014ipa}, and Wilson line configurations are correlated between different values of $x$ \cite{Schenke:2016ksl}.

This additional information can be of great use in the description of the initial state in other collision systems, such as heavy ion collisions performed at the Relativistic Heavy Ion Collider (RHIC) and  at the Large Hadron Collier (LHC). In particular the three dimensional structure of the initial state \cite{Schenke:2016ksl} and the transverse geometry in proton-heavy ion collisions \cite{Mantysaari:2017cni} rely strongly on the detailed configuration of Wilson lines in the incoming target and projectile. Apart from the fundamental interest in the proton and nuclear structure at small $x$, it is thus crucial to understand the Wilson line configurations in protons and nuclei at high energy from electron scattering events in order to constrain the initial states in complex nuclear collisions.

Avoiding approximations like those done in the IPsat or bCGC models and performing explicit small $x$ evolution of a finite size system comes with a variety of problems. One of the issues we will encounter has already appeared when studying the BK evolution of finite size systems \cite{GolecBiernat:2003ym,Ikeda:2004zp,Marquet:2005zf,Berger:2010sh,Berger:2011ew}: The dipole amplitude decreases at large $r$ and eventually vanishes, while in the parametrized models it is always assumed to approach one, at any impact parameter $b$. The behavior in our framework is easy to understand. For increasing $r = |\mathbf{r}|$, it will be more and more likely that both ends of the dipole (the quark and anti-quark) miss the target, and evaluating both Wilson lines in Eq.\,(\ref{eq:dipole_wilsonline_intro}) in the vacuum will lead to $N=0$. 

At $r$ values greater than the inverse pion mass, non-perturbative confinement effects should appear. A simple dipole at such large $r$ is not the right degree of freedom and a more complicated non-perturbative soft contribution, for example described via vector meson dominance \cite{Gribov:1968gs,Brodsky:1969iz,Ritson:1970yu,Sakurai:1972wk} should take over. We will detail this problem further in the main text of this work.

Beyond this important problem, other issues, that can be easily avoided in more ad-hoc calculations of the dipole amplitude, appear. For example, we will have to introduce infrared regulators for both the initial condition of the evolution as well as the JIMWLK kernel, to regulate otherwise appearing Coulomb tails. Again, this is a problem that arises because we cannot deal with confinement effects from first principles. Also, an ultraviolet regulator could be required to modifiy the initial transverse momentum spectrum of gluons, which affects also the $Q^2$-dependence of the cross sections. This regulator is similar to the anomalous dimension used in \cite{Albacete:2004gw}, that modifies the UV behavior of the unintegrated gluon distribution, which in our case is that of the McLerran-Venugopalan model \cite{McLerran:1993ni,McLerran:1997fk} in the initial condition.

The remainder of this paper is organized as follows. First, in Sec.~\ref{sec:jimwlk} we discuss how the small-$x$ evolution of the proton shape is obtained from the JIMWLK evolution. The applied model for the in principle non-perturbative initial condition for this evolution is presented in Sec.~\ref{sec:ic}.
Details of the numerical implementation are given in Sec.~\ref{sec:numerics}. The ovservables of interest in this work, proton structure functions and diffractive cross sections, are presented in Sec.~\ref{sec:xs}. In case of protons without geometric substructure, the results for the structure functions are presented in Sec.~\ref{sec:structurefun} and for the diffractive processes in Sec.~\ref{sec:shape}. Finally, we study the small-$x$ evolution of the proton structure fluctuations in Sec.~\ref{sec:fluctuating_geometry}. Finally, the importance of the possible non-perturbatively large dipoles is studied in detail in Sec.~\ref{sec:large_dipoles}.

\section{High energy evolution from the JIMWLK equation}
\label{sec:jimwlk}
At high energy, the scattering of an electron off a hadronic target (proton or heavier nucleus) is described by the electron emitting a virtual photon, which in turn splits into a quark anti-quark pair. This color dipole then propagates eikonally through the target, meaning that the gluon fields of the target merely rotate the incoming probe's color and transfer transverse momentum. The transverse location of the dipole remains unaffected. The color rotation is mediated by Wilson lines, path ordered exponentials of the color fields along the probes trajectory.

In the color glass condensate framework, the Wilson lines are stochastic variables with an energy (or equivalently $x$ or rapidity $y$) dependent probability distribution. 
This energy dependence is described by the JIMWLK renormalization group equation, which can be written as a functional Fokker-Planck
equation~\cite{Weigert:2000gi}. This in turn can be expressed as a Langevin equation for the Wilson lines $V_\xt$ (in the fundamental representation) themselves~\cite{Blaizot:2002np},
\begin{equation}\label{eq:langevin1}
\frac{\der}{\der y} V_\xt = V_\xt (i t^a) \left[
\int \der^2 \zt\,
\varepsilon_{\xt,\zt}^{ab,i} \; \xi_\zt(y)^b_i  + \sigma_\xt^a 
\right].
\end{equation}

The deterministic drift term is 
\begin{equation}
	\sigma_\xt^a = -i \frac{\as}{2\pi^2} \int \der^2 \zt\, S_{\xt-\zt} \tr \left[ T^a U_\xt^\dagger U_\zt \right],
\end{equation}
with $S_\xt = 1/\xt^2$ and $T^a$ the generators of the adjoint representation. $U_\xt$ are Wilson lines in the adjoint representation.

The random noise is Gaussian and local in coordinates, color, and rapidity with expectation value zero and
\begin{equation}
\label{eq:noice}
\langle \xi_{\xt,i}^a(y) \xi_{\yt,j}^b(y')\rangle = \delta^{ab} \delta^{ij} \delta^{(2)}_{\xt\yt} \delta(y-y').
\end{equation}

The coefficient of the noise in the stochastic term is
\begin{equation}
\label{eq:noicecoef}
 \varepsilon_{\xt,\zt}^{ab,i} = \left(\frac{\as}{\pi}\right)^{1/2}\;
K_{\xt-\zt}^i
\left[1-U_\xt^\dag  U_\zt\right]^{ab},
\end{equation}
where
\begin{equation}
K_\xt^i = \frac{x^i}{\xt^2}.
\end{equation}

When Eq.~\eqref{eq:langevin1} is discretized, the Wilson line at higher rapidity $y+\der y$ is obtained as
\begin{equation}
V_\xt(y+\der y) = V_\xt(y) \exp \left\{ i t^a \int \der^2 \zt\, \varepsilon_{\xt,\zt}^{ab,i} \; \xi_{\zt.i}^b \sqrt{\der y} + \sigma^a_\xt \der y \right\}.
\end{equation}
The delta function in \eqref{eq:noice} is replaced by $\delta(y_m-y_n) \to \delta^{mn}/\der y$ when discretizing the evolution on a lattice.

Following Ref.~\cite{Lappi:2012vw} the drift term can be eliminated, avoiding the necessity to evaluate adjoint Wilson lines, and one rapidity step can be cast into the form
\begin{multline}
\label{eq:wline_evolution_step}
V_\xt(y+\der y) = \exp \left\{ -i \frac{\sqrt{\as \der y}}{\pi} \int \der^2 \zt\, \mathbf{K}_{\xt-\zt} \cdot (V_\zt \boldsymbol{\xi}_\zt V^\dagger_\zt) \right\}\\
\times V_\xt(y) \exp \left\{ i\frac{\sqrt{\as \der y}}{\pi} \int \der^2 \zt\, \mathbf{K}_{\xt-\zt} \cdot \boldsymbol{\xi}_\zt \right\},
\end{multline}
where $\boldsymbol{\xi}_\zt = (\boldsymbol{\xi}_{\zt,1}^a t^a, \boldsymbol{\xi}_{\zt,2}^a t^a)$. 

The long distance Coulomb tails that are encountered when solving rapidity evolution from the JIMWLK equation will give an exponential growth of the cross section with rapidity~ \cite{GolecBiernat:2003ym,Schlichting:2014ipa}, eventually violating the Froissart bound \cite{Kovner:2001bh}. This should be regulated by confinement scale physics, and we regulate the long-distance behavior by following the prescription of Ref.~\cite{Schlichting:2014ipa} and perform the replacement
\begin{equation}
\label{eq:jimwlk_m}
K_\xt^i \to m|\xt| K_1(m|\xt|) \frac{x^i}{\xt^2}.
\end{equation}
Here $K_1$ is the modified Bessel function, and the dimensional parameter $m\sim \lqcd$ will be constrained later.
The replacement (\ref{eq:jimwlk_m}) introduces an exponential suppression at large distances, leaving the short-distance part unmodified.

We will also consider the running coupling effects in our analysis. We adpot the the so called \emph{square root} prescription, in which the coupling constant in Eq.~\eqref{eq:wline_evolution_step} is  moved inside the integral and evaluated as
\begin{equation}
\as(r) =  \frac{12\pi}{(11\nc - 3\nf) \ln \left[ \left(\frac{\mu_0^2}{\lqcd^2} \right)^{1/c} + \left(\frac{4}{r^2 \lqcd^2}\right)^{1/c} \right]^c},
\end{equation}
where $r=|\xt-\zt|$, $\mu_0 =  0.28 \gev$ and $c=0.2$. The scale $\lqcd$ (in coordinate space) will be fixed by fitting the structure function data.

\section{Initial condition for the small-$x$ evolution}
\label{sec:ic}
The initial condition for the JIMWLK evolution, the Wilson lines $V_\xt$ at each point in the transverse plane at the initial rapidity $y$, which corresponds to the initial Bjorken-$x$ $x_0$, is obtained similarly to the IP-Glasma model~\cite{Schenke:2012wb,Schenke:2013dpa}, except that here we do not use any input from the IPsat model. This is preferable, because the parameters in the IPsat model are determined based on a different, parametrized, $x$-evolution.
The color charge densities $\rho^a(\xt)$ in the transverse plane are assumed to be random Gaussian variables with an MV model~\cite{McLerran:1993ni} correlator
\begin{multline}\label{eq:rhoCorrelator}
g^2 \langle \rho^a(x^-, \xt) \rho^b(y^-,\yt) \rangle = \delta^{ab} \delta^{(2)}(\xt-\yt) \delta(x^- - y^-)  \\
\times g^4 \mu^2 T_p\left( \frac{\xt+\yt}{2} \right)
\end{multline}
The color charge density  $g^4\mu^2 T_p(\bt)$  is related to the saturation scale at the given transverse coordinate~\cite{Lappi:2007ku}. We will study both a proton with a Gaussian shape and one that includes additional fluctuations via three randomly positioned hotspots \cite{Mantysaari:2016ykx,Mantysaari:2016jaz}. For the Gaussian proton we have 
\begin{equation}\label{eq:unfactbt}
  T_p(\bt) = \frac{1}{2\pi B_p} e^{-b^2/(2B_p)}\,,
\end{equation}
where $B_p$ controls the proton size. The overall normalization is set by $g^4\mu^2$. 

For the fluctuating proton, we sample the hotspot positions in the transverse plane relative to the origin, $\bti$, from a Gaussian distribution with width $B_{qc}$, assuming a uniform angular distribution. The density profile of each hotspot in the transverse plane is also assumed to be Gaussian
\begin{equation}
T_q(\bt-\bti) = \frac{1}{2\pi B_q} e^{-(\bt-\bti)^2/(2B_q)}\,,
\end{equation}
with width parameter $B_q$.
Thus, the extension to a fluctuating proton corresponds to the replacement
\begin{equation}\label{eq:TqReplace}
T_p(\bt) \rightarrow \frac{1}{N_q} \sum_{i=1}^{N_q}  T_q(\bt-\bti)
\end{equation}
in Eq.~\eqref{eq:unfactbt}. $N_q$ can be interpreted as the number of large $x$ partons, typically chosen to be 3, for the three constituent quarks. In addition, we also allow the saturation scale of each   constituent quark to fluctuate independently, following a log-normal distribution. As discussed in  Ref.~\cite{Mantysaari:2016jaz}, the fluctuations are sampled according to
\begin{equation}
	P(\ln (g^4\mu^2 / \langle g^4\mu^2\rangle) ) = \frac{1}{\sqrt{2\pi}\sigma} \exp\left[ -\frac{\ln^2 (g^4\mu^2/ \langle g^4\mu^2\rangle)}{2\sigma^2} \right].
\end{equation}
The sampled $g^4\mu^2$ values are furher normalized by the expectation value of the distribution $e^{\sigma^2/2}$ in order not to change the average $g^4\mu^2$.

Once the initial color charge densities are set, the Wilson lines at each point in the transverse plane are obtained by solving the Yang-Mills equation as in Ref.~\cite{Schenke:2012wb}. Formally, the classical color field $A^+$ can be written as
\begin{equation}
A^+(x^-, \xt) = -\frac{\rho(x^-, \xt)}{\boldsymbol{\nabla}^2},
\end{equation}
where $\rho = \rho^a t^a$.
This can be written in Fourier space as
\begin{equation}
\label{eq:Aplus_kspace}
A^+(x^-, \kt) = -\frac{\rho(x^-, \kt)}{\kt^2+\tilde m^2}.
\end{equation}
Here, we have introduced an infrared regulator $\tilde m$ that suppresses the non-perturbative long-distance Coulomb  tails. Generally one expects $\tilde m \sim \lqcd$. The mass parameter $\tilde m$ here in momentum space does not necessarily have to be the same as $m$ used in the JIMWLK equation in Eq.~\eqref{eq:jimwlk_m}, which is written in coordinate space, except that we require both of them to be of the order of $\lqcd$. In practice, we use $\tilde m = 0.4\gev$, as it is found in Ref.~\cite{Mantysaari:2016jaz} to produce a Gaussian spectrum for coherent diffractive $J/\Psi$ production, compatible with the experimental data. 

The standard MV model is known to give faster $Q^2$ evolution than what is seen in the HERA structure function data. In dipole model fits using the BK evolved amplitude~\cite{Albacete:2010sy,Lappi:2013zma} this problem is solved by effectively modifying the typical MV-like $r^2$ behavior of the amplitude at small $r$ by introducing an anomalous dimension $\gamma$ that makes the dipole amplitude decrease faster with decreasing $r$. In our setup, to obtain a similar effect we modify the color charge density in momentum space by multiplying Eq.~\eqref{eq:Aplus_kspace} by a suppression factor $e^{-|\kt| v}$:
\begin{equation}
\label{eq:Aplus_kspace_uv}
A^+(x^-, \kt) = -\frac{\rho(x^-, \kt)}{\kt^2+\tilde m^2} e^{- |\kt| v}.
\end{equation}
Here $v$ is a free parameter\footnote{It has been proposed in Ref.~\cite{Dumitru:2011ax} that adding higher order (in color charge) correlators in the action causes the obtained dipole amplitude to have a small-$r$ behavior similar to the one obtained with an anomalous dimension $\gamma > 1$. However, we could not reproduce such behavior in practice.}.

The Wilson lines in coordinate space can finally be obtained by Fourier transforming $A^+(x^-, \kt)$ back to coordinate space and writing \cite{Lappi:2007ku}
\begin{equation}\label{eq:discreteWilson}
V(\xt) = \prod_{i=1}^{N_y} \exp \left( ig A_i^+(\xt) \right),
\end{equation}
where we have also discretized the $x^-$ direction into $N_y$ independent slices. Note that the $A_i^+(\xt)$ in each slice are obtained from a $\rho$ distribution (\ref{eq:rhoCorrelator}) with the amplitude multiplied by $N_y^{-1}$.

\section {Numerical implementation}
\label{sec:numerics}
As described in the previous section, the initial condition at $x=x_0$ is determined by an MV model with a spatially dependent color charge density.
We sample the color charges from a Gaussian determined by (\ref{eq:rhoCorrelator}), compute the Wilson lines \eqref{eq:discreteWilson}, and evolve 
every proton configuration by solving the JIMWLK equation \eqref{eq:wline_evolution_step}. 

As the JIMWLK evolution is implemented as a random walk in color space, different evolutions of the same initial condition will not result in the same configuration at smaller $x$. 
To obtain our final results, we average over 100 initial conditions, each of them evolved only once. 
Calculations are performed on a 2-dimensional $700^2$ lattice, total length being $L=5.12$ fm. We have checked that results have converged and smaller lattice spacings do not modify the results.
The resulting Wilson lines are saved on a grid with rapidity separation $\Delta y \approx 0.2 $, the exact value depending on the chosen $\alpha_s$, and the final results are obtained by interpolating the resulting quantities linearly in $\ln 1/x$.

The free parameters to be determined by the HERA data are the following:
\begin{itemize}
\item $g^4\mu^2$: controls the saturation scale of the proton at $x=x_0$
\item $B_p$: controls the proton size at the initial condition
\item $m$: models confinement effects in the JIMWLK evolution limiting the growth of the proton at long distances
\item constant $\alpha_s$ or $\lqcd$ which sets the scale for the running of $\as$
\end{itemize}
Optional parameters are:
\begin{itemize}
\item $B_{qc}$ and $B_q$: replace $B_p$ in case of a fluctuating proton geometry
\item $v$: ultraviolet damping factor that controls the $Q^2$ evolution speed in the initial condition.
This is correlated with $B_p$, as filtering out the high frequency modes effectively increases the proton size.
\item $\sigma$, which controls the magnitude of the overall saturation scale fluctuations
\end{itemize}

The proton size $B_p$ is chosen to be compatible with the HERA data in the initial condition (we require that the $|t|$ slope of the coherent $J/\Psi$ production in the initial condition is approximately $3.8\gev^{-2}$, see Sec.~\ref{sec:diffraction}). Then, the other parameters are fitted to the HERA structure function data. As we will discuss in detail later, we use the charm reduced cross section to constrain the parameters, as it is not sensitive to the contribution from large dipoles, which are not completely described in our framework. This is because non-perturbative physics becomes important as the dipole size reaches the inverse pion mass. 

As discussed in the introduction, in the dipole picture, cross sections can be expressed in terms of correlators of Wilson lines. In this work, we study proton structure functions and diffractive processes, which are sensitive to the dipole amplitude, which measures the correlation between two Wilson lines
\begin{equation}\label{eq:dipole_wilsonline}
	N\left( \rt, \bt , x  \right) = \left \langle 1 - \frac{1}{\nc} \tr \left( V(\xt,x)V^\dagger(\yt,x) \right) \right \rangle.
\end{equation}
Here, the average $\langle \rangle$ is taken over different color field configurations of the proton, $\bt=(\xt+\yt)/2$ is the impact parameter and $\rt=\xt-\yt$ describes the size and orientation of the color dipole. 

For comparison, we will also show results obtained using the IPsat model parametrization, fitted to HERA data in Ref.~\cite{Rezaeian:2012ji} (see also Ref.~\cite{Mantysaari:2018nng} for a discussion of the large dipole contributions to the structure functions when the IPsat parametrization is applied). In this parametrization, the dipole amplitude is written as
\begin{equation}
N(\rt, \bt, x) = 1-  \exp ( -\rt^2 F(x,\rt^2) T_p(\bt) ),
\end{equation}
 where 
\begin{equation}
F(x, \rt^2) = \frac{\pi^2}{2 \nc} \as(\mu^2) xg(x, \mu^2),
\end{equation}
and $\mu^2 = \mu_0^2 + C/r^2$. The initial condition for the DGLAP evolution \cite{Gribov:1972ri,Altarelli:1977zs,Dokshitzer:1977sg} of the collinear factorization gluon distribution $xg(x, \mu^2)$ and $\mu_0$ are free parameters of the model (where $xg(x, \mu^2=\mu_0^2)$ is a parametrized function of $x$), determined by the fit to HERA data, and $C=4$. The proton density function has the same form as in our JIMWLK calculation, $T_p(\bt) = 1/(4\pi B_p) e^{-\bt^2/(2B_p)}$ with $B_p=4\gev^{-2}$.

\section{Total and exclusive cross sections}
\label{sec:xs}

\subsection{Structure functions}
\label{sec:f2}
Proton structure functions can be written in terms of the virtual photon-proton cross section $\sigma^{\gamma^* p}$ as
\begin{align}
	F_2 &= \frac{Q^2}{4\pi^2 \aem} (\sigma^{\gamma^*p}_L + \sigma^{\gamma^*p}_T), \\
	F_L &= \frac{Q^2}{4\pi^2 \aem} \sigma^{\gamma^*p}_L ,
\end{align}
where $T$ and $L$ refer to transverse and longitudinal polarization of the virtual photon.
The most precise combined HERA data \cite{Abramowicz:1900rp} is released in the form of a reduced cross section
\begin{equation}
	\sigma_r(x,y,Q^2) = F_2(x,Q^2) - \frac{y}{1+(1-y)^2} F_L(x,Q^2),
\end{equation}
where $y=Q^2/(xs)$ is the inelasticity of the $ep$ scattering with center-of-mass energy $\sqrt{s}$.

 In the CGC framework, the virtual photon-proton cross section can be calculated as
\begin{equation}
\label{eq:gamma-xs}
	\sigma^{\gamma^* p}_{L,T} = 2 \sum_f \int \der^2 \bt \der^2 \rt \frac{\der z}{4\pi} \left|\Psi^f_{L,T}(r,z,Q^2)\right|^2 \langle N(\rt, \bt, x) \rangle.
\end{equation}
The target average $\langle \rangle$ refers to averaging over different proton configurations, and the summation is taken over quark flavors $f=\{u,d,s,c\}$. The light quark masses are set to $0.14\gev$, and the charm mass is $1.4\gev$ in this work. Our conclusions are not sensitive to the precise mass values chosen. When evaluating the charm quark contribution we employ the standard kinematical shift~\cite{Tung:2001mv}
\begin{equation}
\tilde x = x\left( 1 + \frac{4m_c^2}{Q^2}\right).
\end{equation}

The virtual photon wave function $\Psi^f_{L,T}$ can be calculated from QED, see e.g.~\cite{Kovchegov:2012mbw}. For the transverse polarization, the squared wave function summed over quark helicities and averaged over photon polarizations reads
\begin{multline}
\label{eq:photon-t}
\left|\Psi^f_{T}(r,z,Q^2)\right|^2 = \frac{2\nc}{\pi} \aem e_f^2 \\ \times \{ [z^2+(1-z)^2\} \varepsilon^2 K_1^2(\varepsilon r)
+ m_f^2 K_0^2(\varepsilon r) \}.
\end{multline}
Similarly, for the longitudinal polarization one obtains
\begin{equation}
\label{eq:photon-l}
\left|\Psi^f_{T}(r,z,Q^2)\right|^2 = \frac{8\nc}{\pi} \aem e_f^2 Q^2 z^2(1-z)^2 K_0^2(\varepsilon r).
\end{equation}
Here $e_f$ is the fractional charge of the quark and $\varepsilon^2 = z(1-z)Q^2 + m_f^2$. $K_0$ and $K_1$ are the modified Bessel functions of the second kind.

\subsection{Diffractive scattering}
\label{sec:diffraction}
Diffractive vector meson production can provide additional insight, in particular on the geometric structure of the target.
For coherent diffractive processes, where the proton stays intact, the cross section can be written as~\cite{Miettinen:1978jb,Kowalski:2006hc}
\begin{equation}
\label{eq:coherent}
\frac{\der \sigma_{T,L}^{\gamma^* p \to V p}}{\der t} = \frac{1}{16\pi} \left| \langle \A_{T,L}^{\gamma^* p \to V p}(\xpom,Q^2,\boldsymbol{\Delta}) \rangle \right|^2,
\end{equation}
where $\A_{T,L}^{\gamma^* p \to V p}(\xpom,Q^2,\boldsymbol{\Delta})$ is the scattering amplitude. The indices $T,L$ indicate the photon polarization. 
In case of diffractive events, we will only consider photoproduction ($Q^2 = 0$). Consequently, only the transverse polarization will appear.

The incoherent cross section can be written as the variance~\cite{Miettinen:1978jb} (see also Refs.~\cite{Frankfurt:1993qi,Frankfurt:2008vi,Caldwell:2009ke,Lappi:2010dd, Mantysaari:2016ykx, Mantysaari:2016jaz}):
\begin{align}\label{eq:incoherent}
\frac{\der \sigma_{T,L}^{\gamma^* p \to V p^*}}{\der t} = \frac{1}{16\pi} &\left( \left\langle \left| \A_{T,L}^{\gamma^* p \to V p}(\xpom,Q^2,\boldsymbol{\Delta})  \right|^2 \right\rangle \right. \notag\\ 
& ~~~ - \left. \left| \langle \A_{T,L}^{\gamma^* p \to V p}(\xpom,Q^2,\boldsymbol{\Delta}) \rangle \right|^2 \right)\,.
\end{align}

\begin{figure*}[htb]
\centering
		\includegraphics[width=0.7\textwidth]{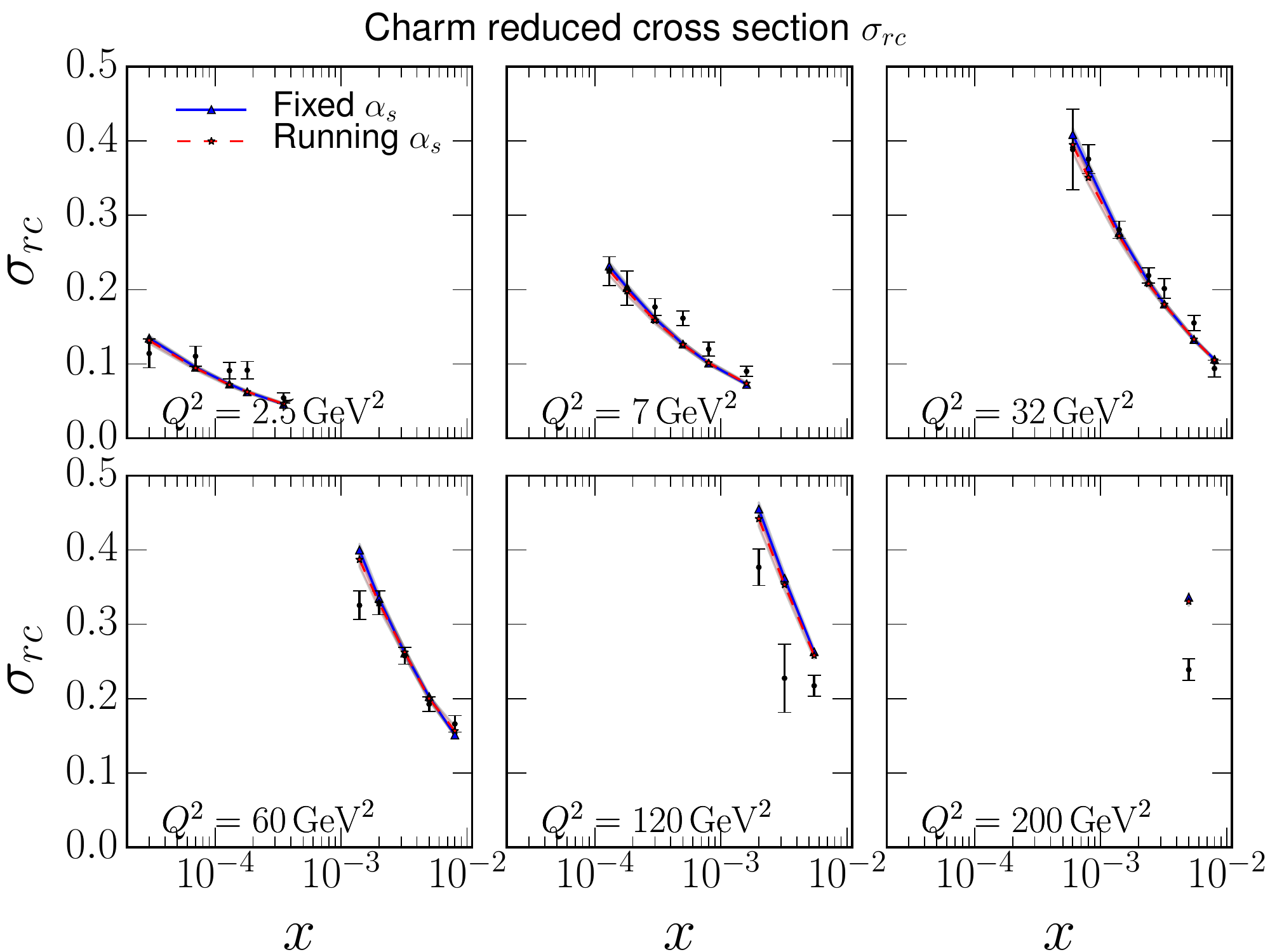} 
				\caption{Comparison to the HERA charm structure function data~\cite{H1:2018flt} with both fixed ($\as=0.21$) and running coupling ($\lqcd=0.09\gev$). The band represents the statistical uncertainty of the calculation. The fit quality is approximately $\chi^2/N=4.3$ (fixed coupling) or $\chi^2/N=3.9$ (running coupling).}
		\label{fig:sigmar_charm}
\end{figure*}

 \begin{figure*}[htb]
\centering
		\includegraphics[width=0.7\textwidth]{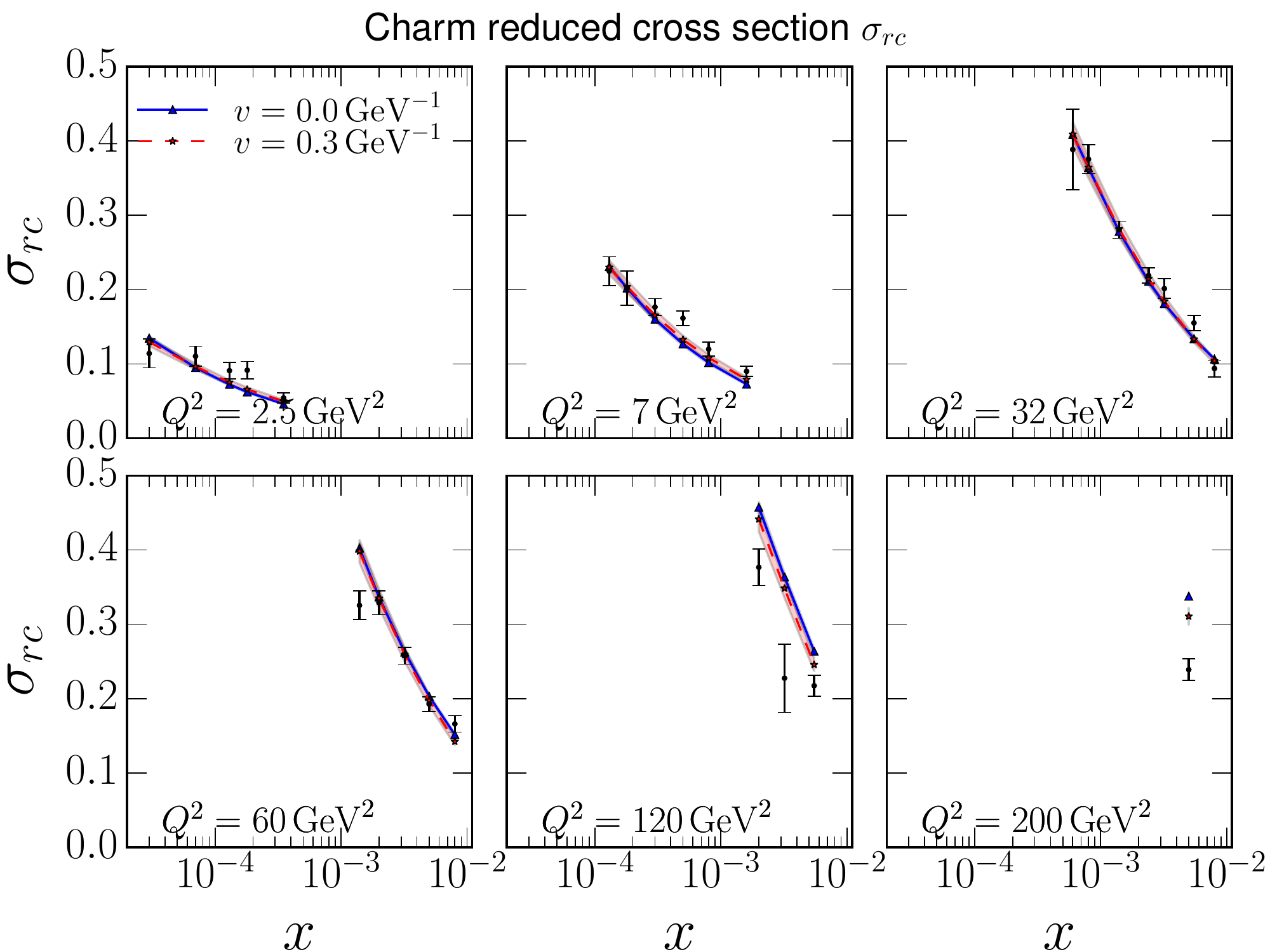} 
				\caption{Description of the charm reduced cross section using the unmodified MV model ($v=0, \chi^2/N =4.3$)  and with an ultraviolet damping in the initial condition ($v=0.3\gev^{-1}, \chi^2/N=2.7$).
				}
		\label{fig:sigmar_charm_uv}
\end{figure*}

\begin{figure*}[htb]
\centering
		\includegraphics[width=0.7\textwidth]{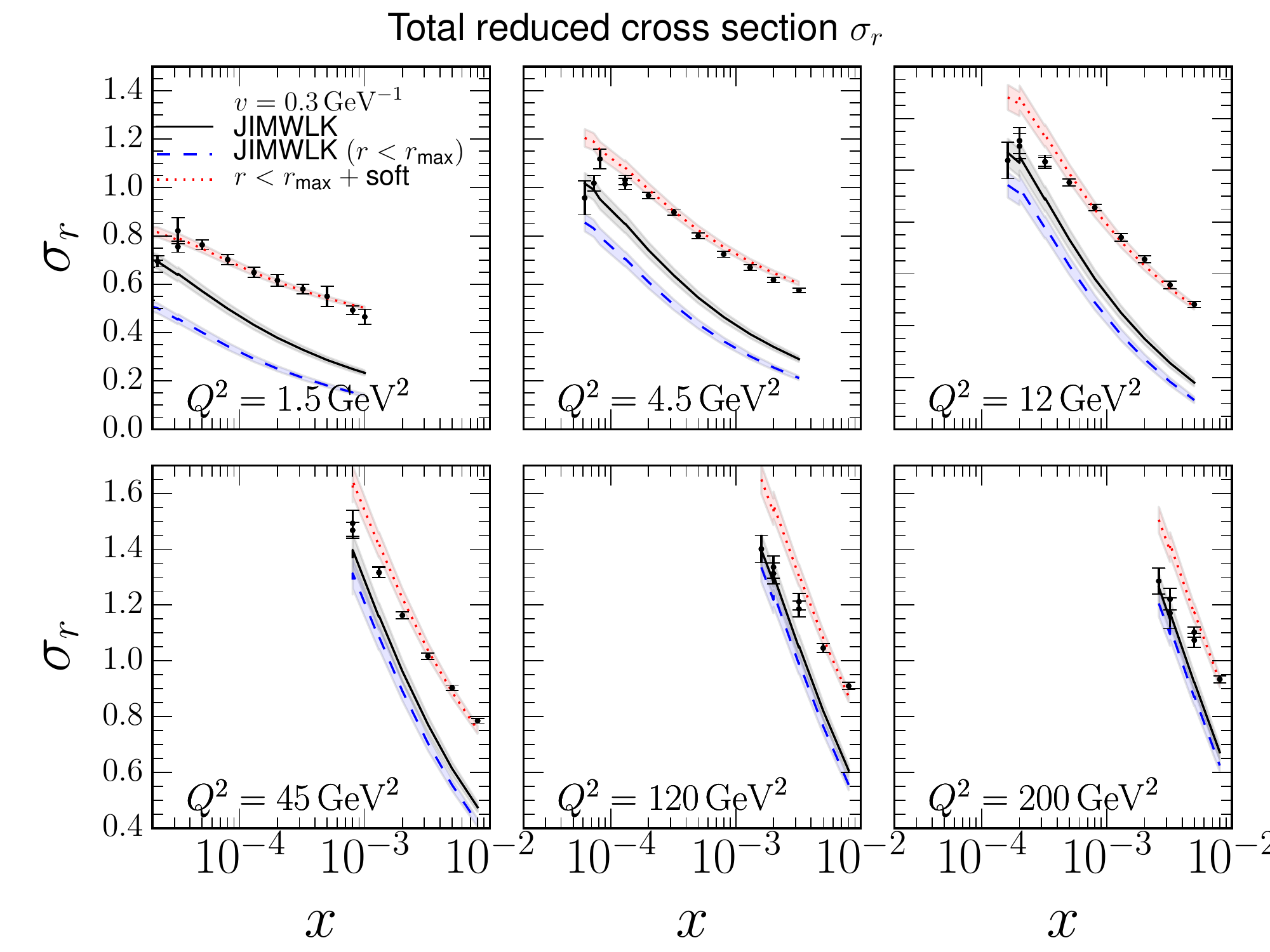} 
				\caption{HERA total reduced cross section data~\cite{Abramowicz:1900rp} compared with the result from the JIMWLK evolution with ultraviolet damping at fixed coupling. The band represents the statistical uncertainty of the calculation. The solid line shows the full fixed coupling JIMWLK result. The dashed line includes only contributions from dipoles smaller than $r_\text{soft}=0.4\fm$, and the dotted line includes additional soft component (see text for details). } 
						\label{fig:sigmar_total}
\end{figure*}

The diffractive vector meson production scattering amplitude can be written as~\cite{Kowalski:2006hc},
\begin{multline}
\label{eq:diff_amp}
 \A^{\gamma^* p \to V p}_{T,L}(\xpom,Q^2, \boldsymbol{\Delta}) = i\int \der^2 \rt \int \der^2 \bt \int \frac{\der z}{4\pi}  \\ 
 \times (\Psi^*\Psi_V)_{T,L}(Q^2, \rt,z) \\
 \times e^{-i[\bt - (1-z)\rt]\cdot \boldsymbol{\Delta}}  2N(\rt,\bt,\xpom).
\end{multline}
Here the transverse momentum transfer is $\Deltat=(P'-P)_\perp \approx \sqrt{-t}$. This equation can be interpreted as follows: First, the incoming virtual photon fluctuates into a quark-antiquark dipole with transverse separation $\rt$, the quark carrying the momentum fraction $z$. This splitting is described by the same virtual photon wave function $\Psi_{T,L}$ that is used when calculating the proton structure function. The elastic scattering amplitude for the dipole to scatter off the target is $N(\rt,\bt,\xpom)$. Finally, the vector meson is formed, and the $q\bar q \to V$ formation is described by the vector meson wave function $\Psi_V$. Note that $\Psi_V$ is a significant source of uncertainty in our calculations. In this work we use the Boosted Gaussian wave function parametrization from Ref.~\cite{Kowalski:2006hc} (see Ref.~\cite{Chen:2016dlk} for a new development of more rigorous calculations of the vector meson wave functions).

When calculating the cross section, we also take into account the so called skewedness correction, which originates from the fact that in the dilute limit, the two exchanged gluons carry significantly different longitudinal momentum fractions~\cite{Martin:1997wy,Shuvaev:1999ce, Martin:1999wb}. We apply this by calculating the effect of the skewedness correction in the IPsat model at given $W$ (and use the fact that the correction is approximately independent of $t$), and assume that the effect is the same in the formulation using the MV model and JIMWLK evolution.

For $J/\Psi$ production, the structure of the target is probed at the scale
\begin{equation}
\xpom = \frac{Q^2+M_{J/\Psi}^2-t}{Q^2+W^2-m_N^2},
\end{equation}
which can be interpreted as the longitudinal momentum fraction of the proton carried by the color-neutral ``pomeron''. Here $W$ is the center-of-mass energy in the photon-proton scattering, and in this work we neglect the $t$ dependence of $\xpom$, as $|t| \ll M_{J/\Psi}^2$. 

The cross section is Fourier transformed into momentum space, and the transverse momentum transfer $\Deltat$ is the Fourier conjugate to $\bt - (1-z)\rt$. Note that this means that a detailed knowledge of the impact parameter profile is needed, in contrast to the calculations of the proton structure functions where the impact parameter integral  mainly affects the overall normalization.  This makes diffractive scattering a sensitive probe of the proton geometry and, via Eq. \eqref{eq:incoherent}, its fluctuations.

\section{Description of the proton structure function data}
\label{sec:structurefun}

The proton structure functions and the reduced cross sections have been measured very accurately by the HERA experiments H1 and ZEUS~\cite{Aaron:2009aa,Abramowicz:1900rp,Abramowicz:2015mha,H1:2018flt} over a wide range of $x$ and $Q^2$. The experimental data is released separately for the total reduced cross section, and for the charm contribution to it. 
In case of transverse polarization, the total photon-proton cross section includes the so called ``aligned jet'' contribution where the momentum fraction $z\sim 1$ or $z\sim 0$.
In this limit, the dipole sizes are limited by $1/m_f^2$ instead of $1/Q^2$ (see Eq.~\eqref{eq:photon-t}), and thus at all $Q^2$ there is a significant contribution to $F_2$ from possibly non-perturbatively large dipoles  which would be sensitive to the confinement scale physics. 
As these effects are not included in our analysis (see also \cite{Berger:2011ew}), we prefer to first fit the charm structure function data (along with the proton size constraint from diffractive $J/\Psi$ production), where the large charm mass suppresses confinement scale effects at all $Q^2$.
Note that in case of $F_L$, where only longitudinally polarized photons contribute, the aligned jet contribution is absent, as the endpoints $z\to 0,1$ are suppressed by a factor $z^2(1-z)^2$, see Eq.~\eqref{eq:photon-l}.

The slope of the $|t|$ dependent coherent diffractive cross section for exclusive $J/\Psi$ production is used to fix $B_p$ in the initial condition ($x=x_0=0.01$), fixing the proton size at that value of $x$. Then, by varying $g^4\mu^2$ and $\as$ (at fixed coupling) or $\lqcd$ (with running coupling), and calculating the Wilson lines at smaller $x$ by solving the JIMWLK equation, we find the parametrization that gives approximately the best possible $\chi^2$ when comparing with the HERA charm structure function data from Ref.~\cite{H1:2018flt}.

Note that as the total photon-proton cross section is proportional to both the (squared) saturation scale and the geometric size of the proton, 
the evolution speed is controlled by both the coupling constant $\as$ and the infrared regulator $m$ in the JIMWLK equation that controls the growth of the proton size.
 Thus, the HERA reduced cross section data is not enough to determine uniquely these two parameters. The fit quality remains excellent if $m$ is increased (decreased), if one also increases (decreases) the strong coupling constant. This is demonstrated in Appendix~\ref{appendix:mdep}, where we show that varying $m$ by $50\%$ does not affect the quality of the fit when $\as$ is adjusted (it is strongly correlated with $m$). In this work, unless otherwise noted we use $m=0.2\gev$. However, as $m$ controls the evolution of the proton geometry at long distances, it has a large effect on the small-$|t|$ part of the coherent vector meson production spectra, where the structure at long distances is probed. This is demonstrated in Appendix~\ref{appendix:mdep}. This behavior could make it possible to constrain both $\alpha_s$ and $m$ individually.

Comparison to the HERA combined charm structure function data~\cite{H1:2018flt} with our results for the optimal parameters is shown in Fig.\,\ref{fig:sigmar_charm} using the MV model initial condition without proton geometry fluctuations.
The description of the data is equally good with both fixed and running coupling ($\chi^2/N \approx 4$).
The too fast $Q^2$ evolution of the MV model discussed in Sec.~\ref{sec:ic} is clearly visible\footnote{Recent next-to-leading order calculations~\cite{Beuf:2017bpd,Ducloue:2017ftk} suggest that the higher order corrections may also slow down the $Q^2$ evolution speed.}, but at moderate $Q^2$ the agreement with the HERA data is good. The IPsat results are not shown, but they would be on top of the HERA data~\cite{Rezaeian:2012ji}.

Fig.\,\ref{fig:sigmar_charm_uv} shows the effect of including a UV regulator $v=0.3\,{\rm GeV}^{-1}$ according to Eq.\,\eqref{eq:Aplus_kspace_uv}. One can see an improvement of the $Q^2$ evolution speed compared to the experimental data. More significant than what is visible by eye is the improvement of the fit, quantified by $\chi^2/N$ which becomes $\approx 2.7$, when including the UV damping. In general, we find that our fit prefers large values for the ultraviolet regulator $v$, with increasing $v$ also decreasing the extracted value for $\as$. We were not able to find a single $v$ value preferred by the fit, and thus fix it to $v=0.3\,\gev^{-1}$ when showing results with ultraviolet damping. 

The model parameters used to describe the charm production data are shown in Table~\ref{table:round_parameters}. Note that as already discussed, the ultraviolet damping factor $e^{-|\kt| v}$ in Eq.~\eqref{eq:Aplus_kspace_uv} effectively makes the proton larger by filtering out the high frequency modes, which is compensated by a more sharply peaked initial color charge distribution in the MV model at the origin ($B_p$ is much smaller than in the case $v=0$). Similarly, as this suppression factor clearly reduces the overall color charge density, this effect is compensated by a larger $g^4\mu^2$ in the initial condition. The physical interpretation of the parameters, for example as color charge density and proton size, are somewhat obscured in case the ultraviolet regulator is employed. Note that in both parametrizations the resulting saturation scales extracted from the dipole amplitude are very similar.

\begin{table}[tp]
\begin{center}
\begin{tabular}{c|c|c|c|c}
$\chi^2/N$ & $v$ $[\gev^{-1}]$ & $g^4\mu^2$ $[\gev^2]$ & $B_p$  $[\gev^{-2}]$ &  $\alpha_s$  \\
 \hline
4.3 & 0.0 & 0.8 & 2.1 &  0.21  \\
2.7 & 0.3 & 4.25 & 1.2 & 0.18
 \end{tabular}
\end{center}
\caption{Parametrizations used to describe the HERA charm reduced cross section data~\cite{H1:2018flt} without including geometric fluctuations, with the proton size constrained by the coherent $J/\Psi$ spectra. The infrared regulator in the JIMWLK evolution is $m=0.2\gev$. Note that with nonzero ultraviolet damping $v>0$, the physical interpretation of the parameters is somewhat obscured.}
\label{table:round_parameters}
\end{table}%

Using the parametrization constrained by the charm data (and the $|t|$-slope of the exclusive $J/\Psi$ production), we next calculate the total reduced cross section including the light quarks in addition to charm.  The results are shown by the solid lines in Fig.~\ref{fig:sigmar_total}, where we find that especially at $x$ values close to the initial condition the total photon-proton cross section is significantly underestimated in our calculation. Again, the IPsat result is not shown, but the description of the data would be almost perfect ($\chi^2/N\approx 1.2$).
The \emph{soft} contribution, and the contribution from dipoles smaller than $r_\text{soft}=0.4\fm$, are discussed later in this section.

In the IPsat model fits one obtains a good simultaneous description of both charm and total structure function data. As the charm data is well described also by our calculation, the resulting dipole amplitudes must thus be comparable at small dipole sizes. The large differences in the total reduced cross section can then be understood by analyzing the large dipole contributions. In our case, the dipole scattering amplitude vanishes when the quarks miss the proton, which heavily suppresses contributions from dipoles larger than the proton size (note that the gluonic transverse root mean square radius of the proton is quite small, $r_\text{rms}=\sqrt{2B_p} \approx 0.55\fm$ at the initial condition). The proton growth towards smaller $x$ allows finite contributions from larger dipoles, which manifests itself in a better description of the small-$x$ structure function data. On the other hand, this effect could also lead to an artificially large $x$-evolution speed. 

In the IPsat or BK fits~\cite{Albacete:2010sy,Rezaeian:2012ji,Lappi:2013zma}) the dipole amplitude goes to unity at large dipoles indepedently of the impact parameter. In particular in the IPsat parametrization the total dipole-proton cross section scales like $\sim \ln r$ at large dipole sizes $r$, in sharp contrast to the behavior obtained in our framework.
Effectively, this means that in the IPsat fits one models the non-perturbative confinement scale physics by imposing a requirement that large dipoles scatter with probability $1$.

\begin{figure}[tb]
\centering
		\includegraphics[width=0.5\textwidth]{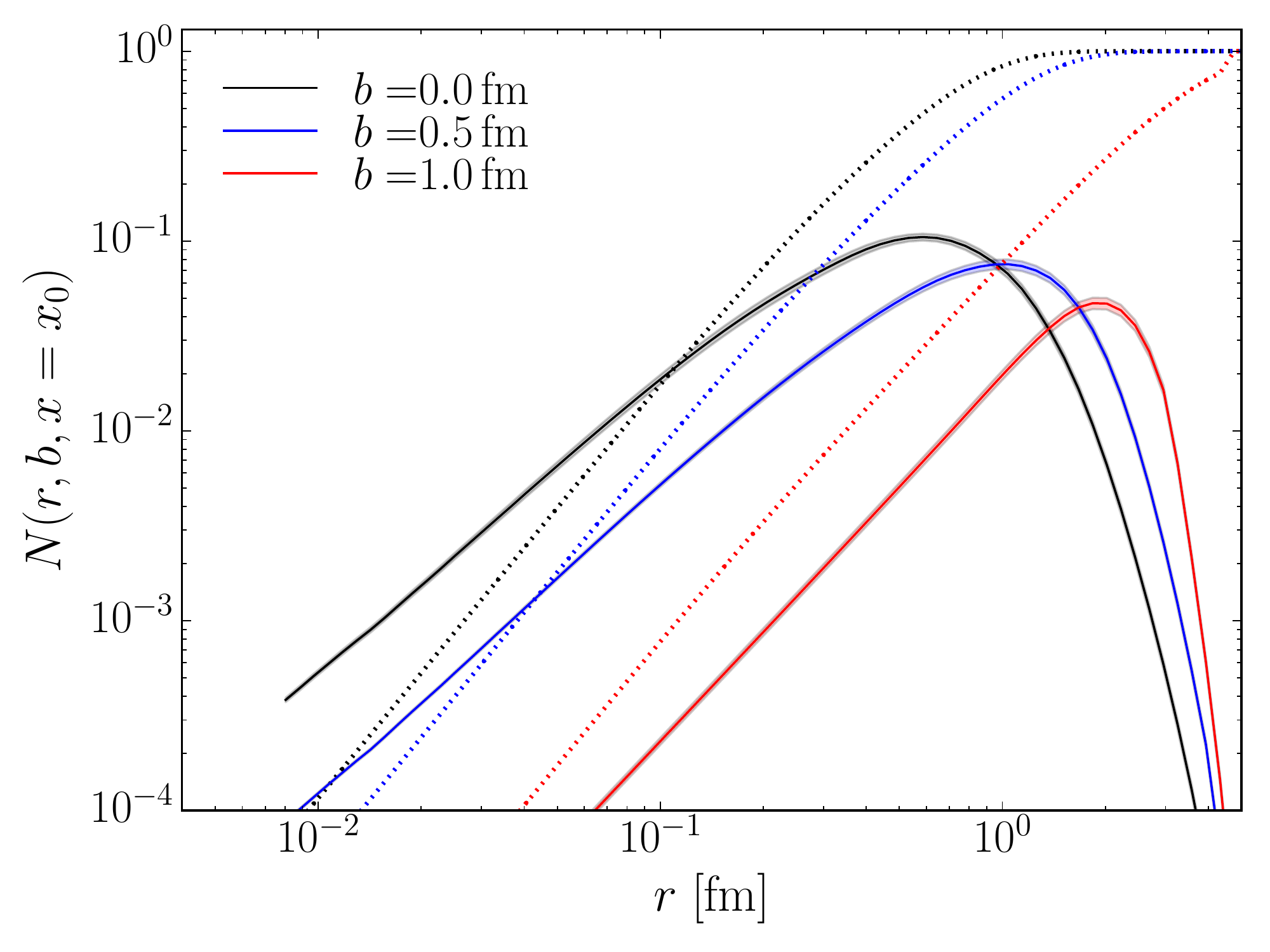} 
				\caption{Dipole ampltiude at the MV model initial condition ($v=0$) at different impact parameters. Dotted lines show comparison to the IP-sat model from Ref.~\cite{Rezaeian:2012ji}. The dipole ampltiude is averaged over an impact parameter range from $b$ to $b+0.05\fm$.  }
		\label{fig:dipole_amplitude_x0}
\end{figure}

To make the above discussion more transparent, we show in Fig.~\ref{fig:dipole_amplitude_x0} the dipole amplitudes in our framework in comparison to those in the IPsat model \cite{Rezaeian:2012ji} in the initial condition for different impact parameters. First, the striking difference between the models at large $r$ is obvious. While the IPsat model always assumes $N\rightarrow 1$ as $r \rightarrow \infty$, the dipole amplitude in our calculation goes to zero at large $r$. Furthermore, dipole amplitudes are found to grow more slowly with the dipole size $r$ than in the IPsat parametrization. This can be understood, as in the IPsat  the probed saturation scale remains approximately constant when the impact parameter is fixed, which is somewhat unphysical. In contrast, in our calculation, when the separation between the quarks increases, at zero impact parameter both of them move to less dense regions of the proton, and the effective saturation scale decreases.

\begin{figure}[tb]
\centering
		\includegraphics[width=0.5\textwidth]{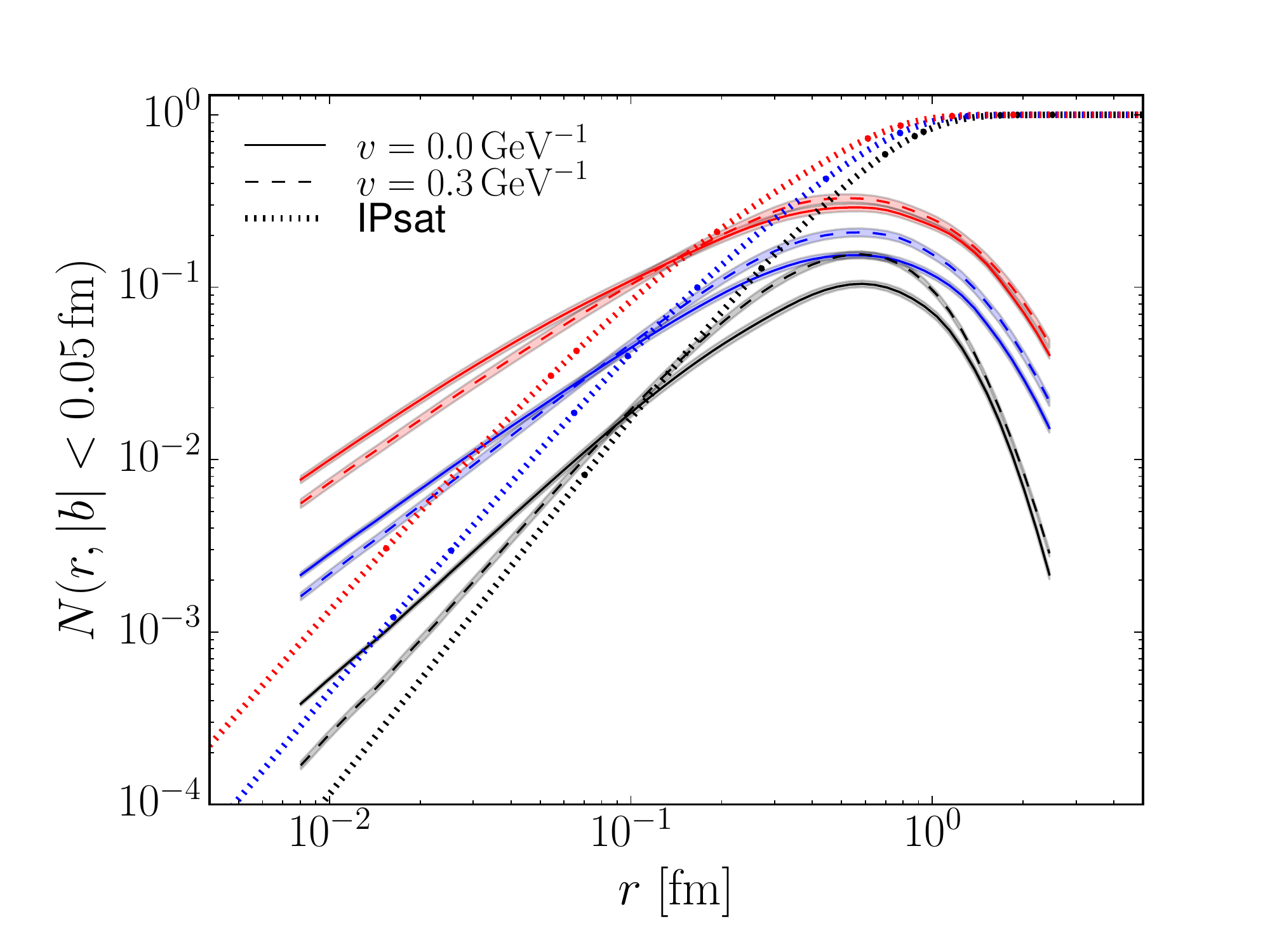} 
				\caption{Dipole amplitude at central impact parameter in the MV model and with UV damping in the initial condition compared with the IPsat result (dotted lines). The rapidities are $y=0.0$, $y=2.4$ and $y=4.8$ (from right to left).  }
		\label{fig:amplitude_uv}
\end{figure}

\begin{figure}[tb]
\centering
		\includegraphics[width=0.5\textwidth]{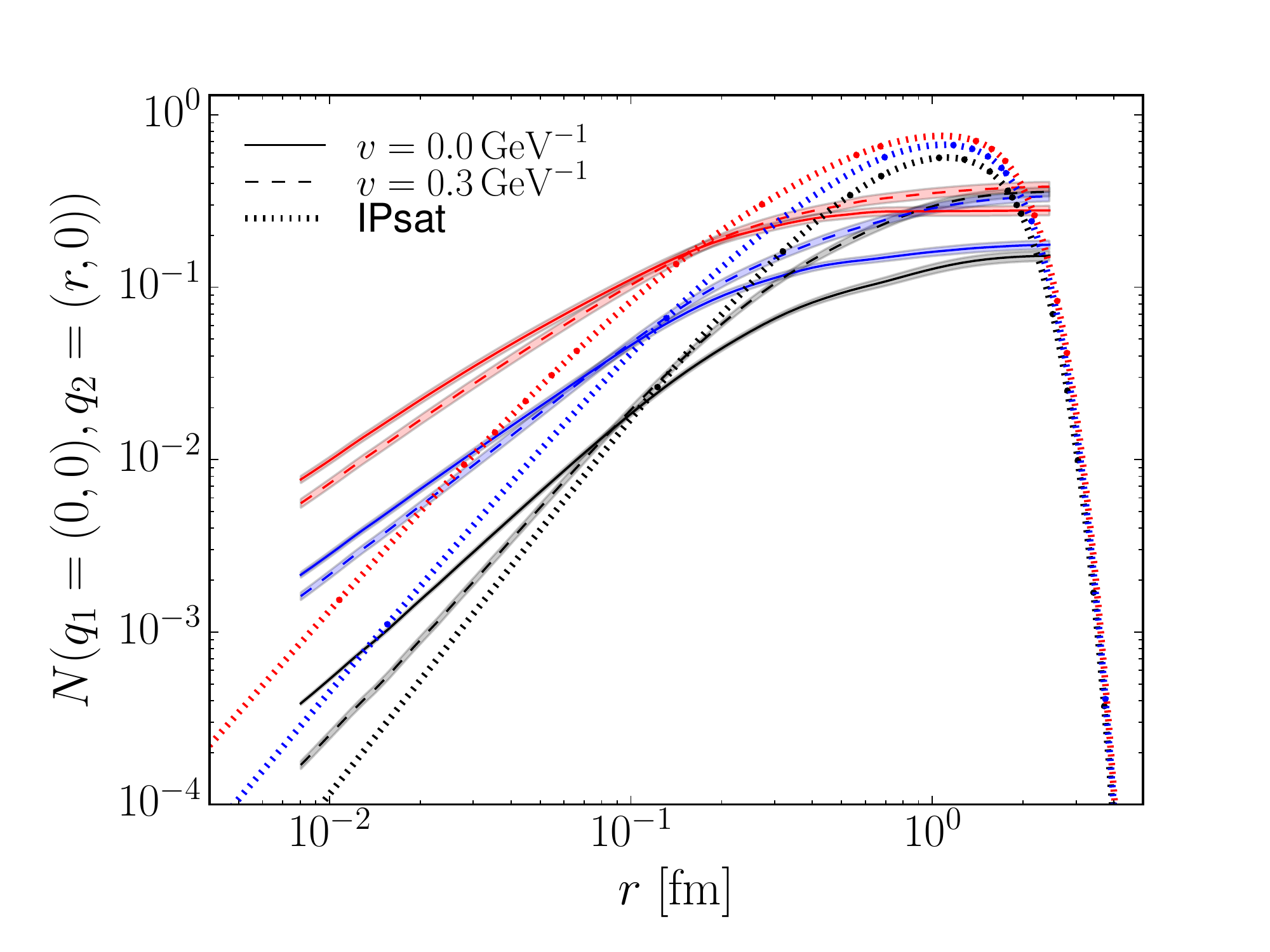} 
				\caption{Dipole amplitude with one quark at the center, the other at distance $r$. The evolution rapidities are $y=0$, $y=2.4$ and $y=4.8$ (from right to left). IPsat comparisons are calculated with $b=r/2$ and shown as dotted lines.}
		\label{fig:amplitude_uv_q1_0}
\end{figure}

 The rapidity evolution of the dipole amplitude is shown in Figs.\,\ref{fig:amplitude_uv} and \ref{fig:amplitude_uv_q1_0}, where we again compare to the IPsat model. The resulting amplitude with and without UV damping in the initial condition are shown in Fig.\,\ref{fig:amplitude_uv} for the center of the dipole located at the center of the proton, and in Fig.\,\ref{fig:amplitude_uv_q1_0} for one end of the dipole located at the center of the proton. In both cases we can see that inclusion of the UV damping introduces an effective anomalous dimension and at small $r$ the dipole is similar to that in the IPsat parametrization. The JIMWLK rapidity evolution effectively reduces the anomalous dimension (which also happens with BK evolution \cite{Albacete:2007yr}), and deviations from IPsat increase.

If one end of the dipole remains in the center of the proton, the dipole amplitude always increases with increasing dipole size as shown in Fig.\,\ref{fig:amplitude_uv_q1_0}, in contrast to the $b\approx 0$ case. This can be understood, because the dipole amplitude will go to zero only when both Wilson lines are the vacuum ones. The effect of dropping dipole amplitude at large $r$ for IPsat in the case where  one end of the dipole is held fixed occurs because in that case increasing $r$ leads to increasing $b$ (the central point between the dipole ends), and the probed saturation scale is exponentially suppressed as $Q_s^2 \sim e^{-b^2/{2B_p}}$.

Within a model using impact parameter dependent BK evolution \cite{Berger:2011ew}, it was found that a numerically significant non-perturbative ``soft'' contribution to the total cross section must be included to describe the experimental structure function data. In order to quantify how large we expect the soft contribution to be, we follow the approach of Ref.~\cite{Berger:2011ew} and separate contributions to the structure functions into perturbative and soft parts. The soft contribution to the structure functions is calculated as
\begin{align}
	F_2^\text{soft} &= \frac{Q^2}{2\pi \aem} \sigma_0 \int_{r_\text{soft}} \der r \, r \, \int \der z (|\Psi_L|^2 + |\Psi_T|^2) \\
	F_L^\text{soft} &= \frac{Q^2}{2\pi \aem} \sigma_0 \int_{r_\text{soft}} \der r \, r \, \int \der z |\Psi_L|^2,
\end{align}
and corresponding perturbative contributions are computed as discussed above and imposing an upper limit $r < r_\text{soft}$ for the dipole sizes. The total structure functions (and finally the total reduced cross section) are obtained as a sum of perturbative and soft components. For the charm production the soft component has a negligible effect.

The resulting reduced cross section (that includes the soft contribution) is also show in 
Fig~\ref{fig:sigmar_total}. We use $r_\text{soft}=0.4\fm$ as a scale to separate soft and hard physics, and the proton transverse area is fixed to $\sigma_0/2 = 13.6$ mb in order to get a good description of the HERA data in the lowest $Q^2$ bin. The parameter values are close to the values used in Ref.~\cite{Berger:2011ew} ($0.56 \fm$ and 14.6 mb, respectively). When the soft component is included, the description of the HERA data is good, except at very high $Q^2$ where too fast $Q^2$ dependence is again observed similar to the case of the charm structure function. 
We note that in principle the soft contribution is expected to also be $x$ dependent, but there is currently no way to compute this $x$ dependence from first priciples, as it is done for the perturbative contribution. The dotted line in Fig.\,\ref{fig:sigmar_total} represents the perturbative contribution for $r<r_\text{soft}$, to which the soft contribution is added.

\section{Exclusive J/$\Psi$ production}
\label{sec:shape}

\begin{figure}[tb]
\centering
		\includegraphics[width=0.5\textwidth]{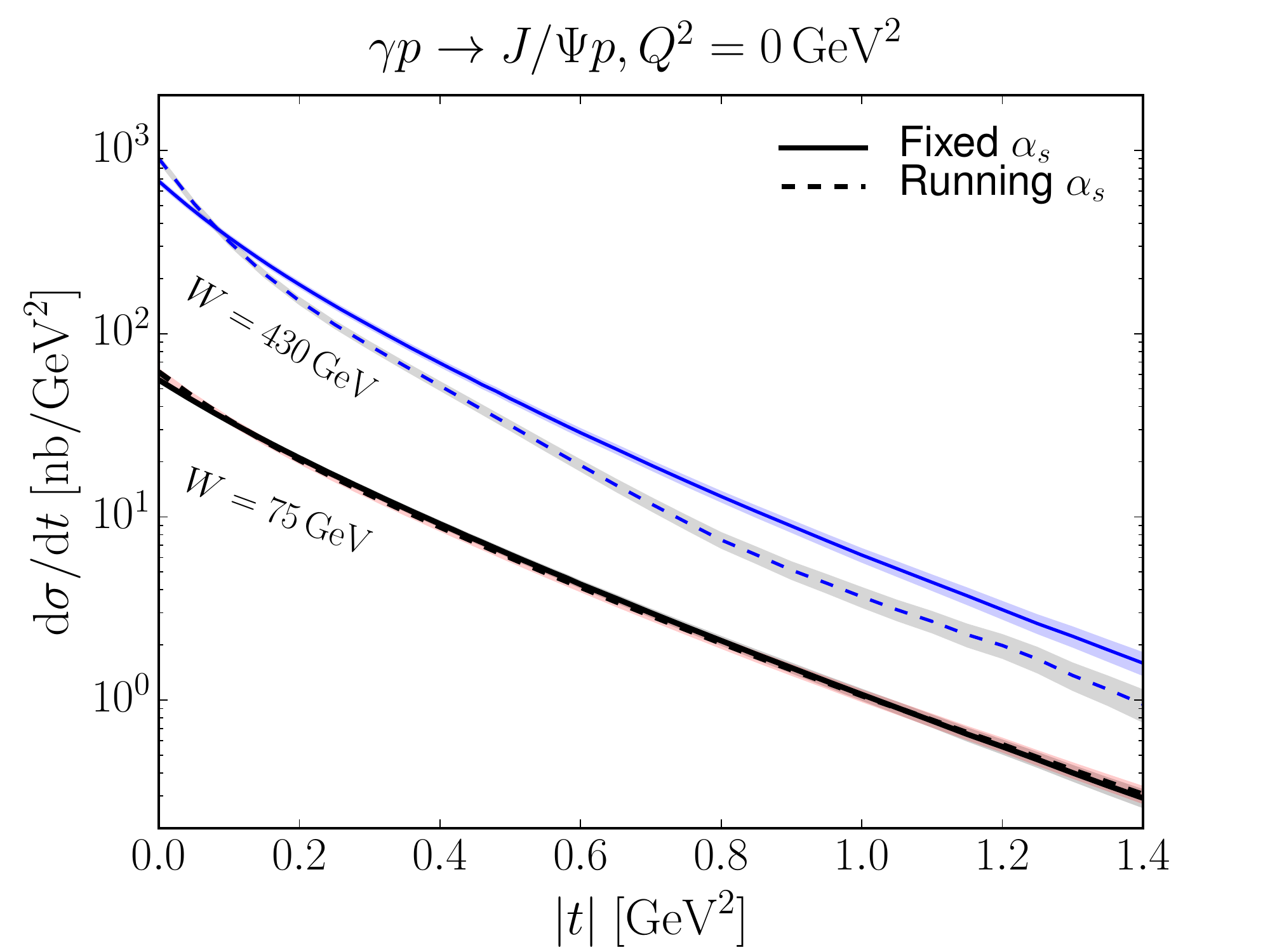} 
				\caption{Coherent diffractive $J/\Psi$ production cross section as a function of squared momentum transfer at two different center-of-mass energies, $W=75\gev$ (lower black lines) and $W=440\gev$ (upper blue lines). Initial condition is MV model ($v=0$).}
		\label{fig:jpsi_t}
\end{figure}

The coherent diffractive $J/\Psi$ production cross section as a function of squared momentum transfer $|t|$ is shown in Fig.~\ref{fig:jpsi_t}. The results at both fixed and running coupling are shown at $W=75$\,\gev (which can be compared with the H1 data~\cite{Alexa:2013xxa}, but again due to the lack of the large dipole contributions the calculation would underestimate the data as discussed in more detail later) and at high energies $W=440\gev$. Note that these energies correspond to evolution of $1.8$ and $5.3$ units of rapidity  from the initial condition, respectively.
After a few units of rapidity evolution the results obtained with fixed and running coupling start to deviate. 
With running coupling, the low-$|t|$ part of the cross section is enhanced compared to the case of fixed coupling. The reason for this is that the running coupling increases the evolution speed of the long-distance modes relative to the short-distance ones, and the low-$|t|$ part is sensitive to long-distance physics ($\sqrt{-t}$ is  Fourier conjugate to the impact parameter). 

 \begin{figure}[tb]
\centering
		\includegraphics[width=0.5\textwidth]{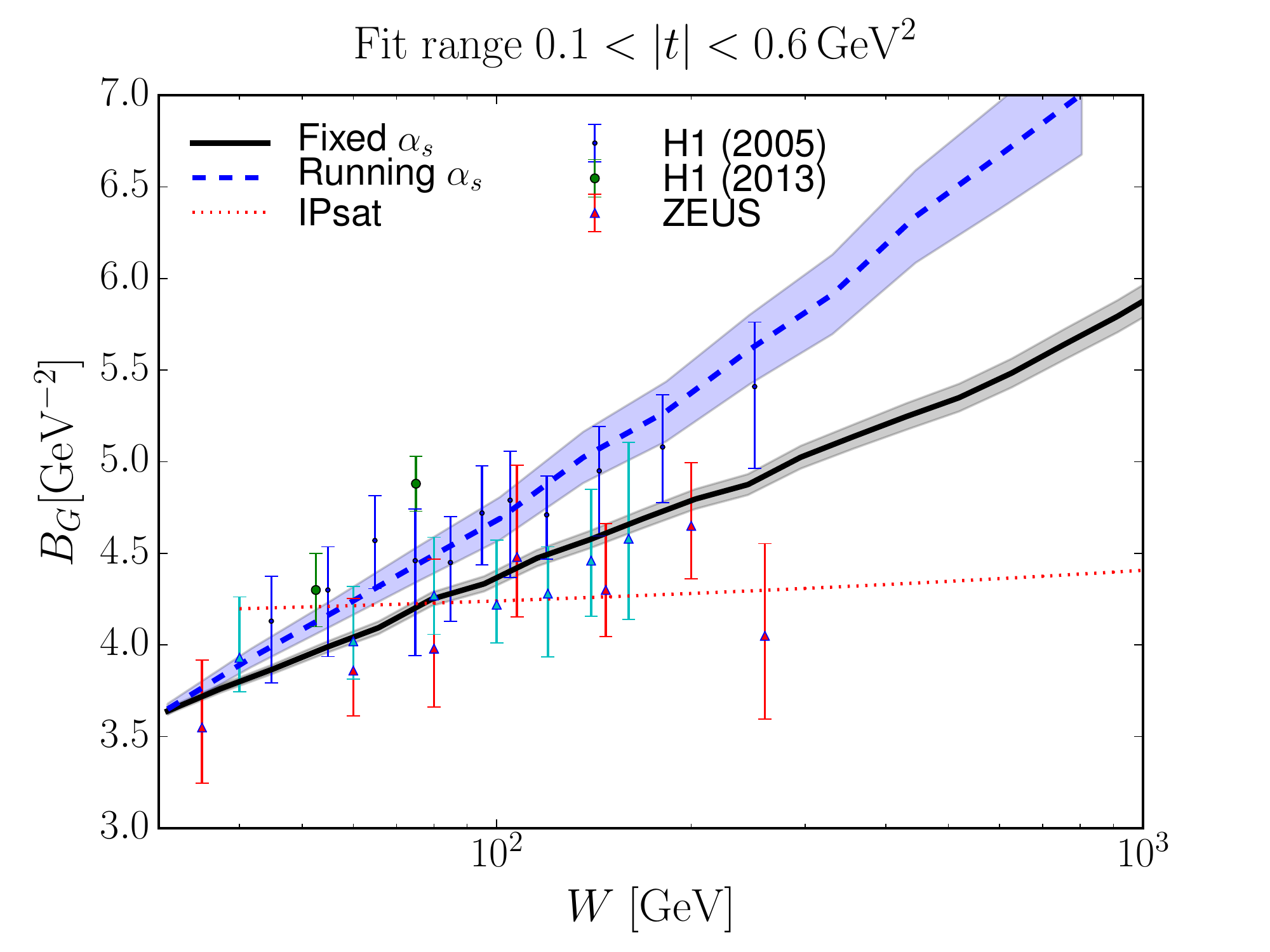} 
				\caption{Diffractive slope $B_G$ from JIMWLK evolution with fixed (solid) and running (dashed) coupling, compared to the result from IPsat (dash-dotted) and experimental data from H1~\cite{Alexa:2013xxa} and ZEUS~\cite{Chekanov:2002xi}. $B_G$ is extracted by fitting the $|t|$ spectrum with an exponential in the range $0.1 < |t| < 0.6\,{\rm GeV}^2$. The initial condition is the MV model ($v=0$).
				}
		\label{fig:proton_size_diffraction}
\end{figure}

\begin{figure*}[tb]

\centering
    	\begin{minipage}{.48\textwidth}
        \centering
		\includegraphics[width=\textwidth]{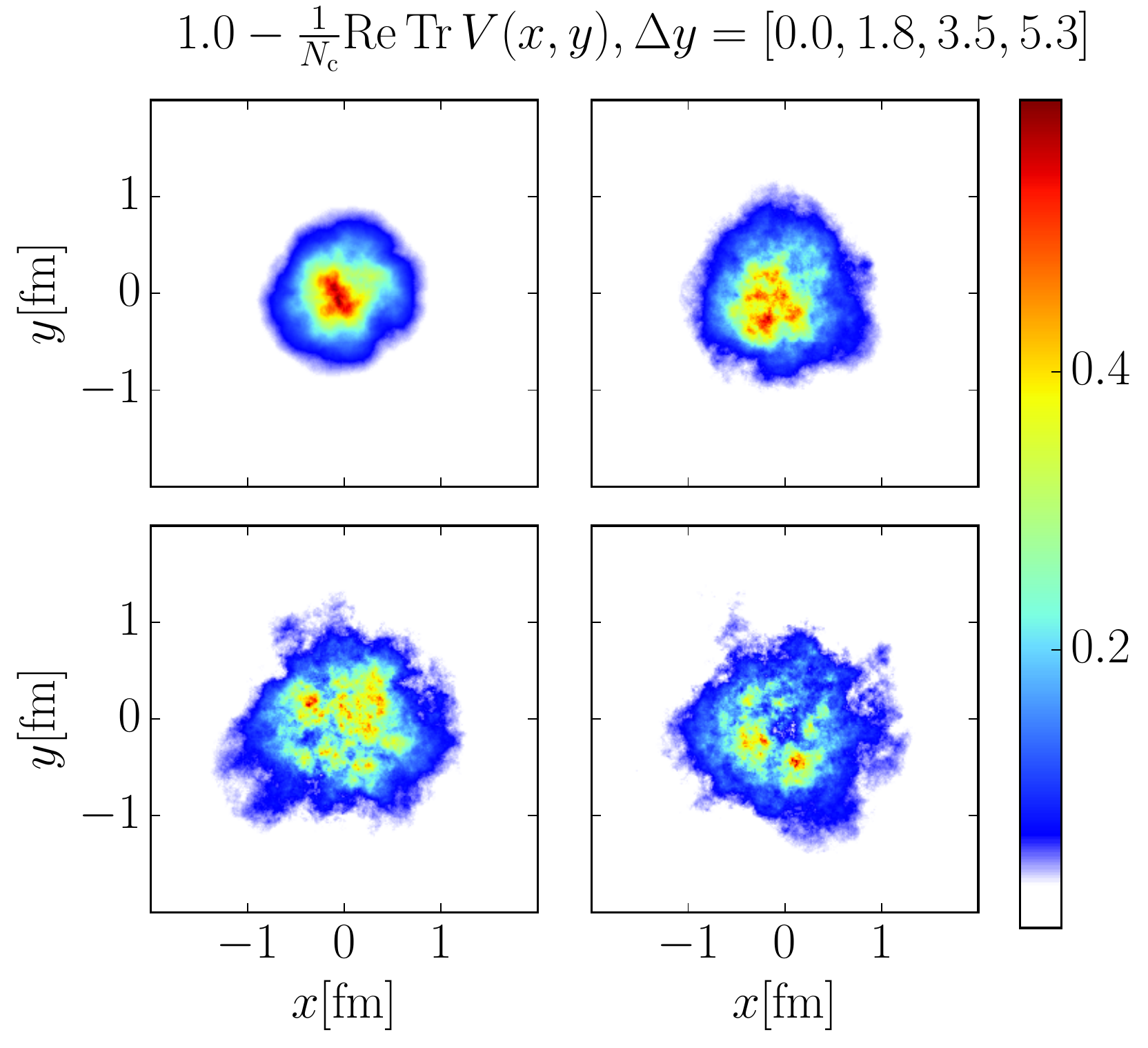} 
				\caption{Example of the proton density profile (illustrated as a trace of the Wilson line) evolution over $5.3$ units of rapidity. The initial condition is MV model ($v=0$).}
		\label{fig:tr_v_ic}
	\end{minipage}
	\quad
	\begin{minipage}{0.48\textwidth}
	\centering
		\includegraphics[width=\textwidth]{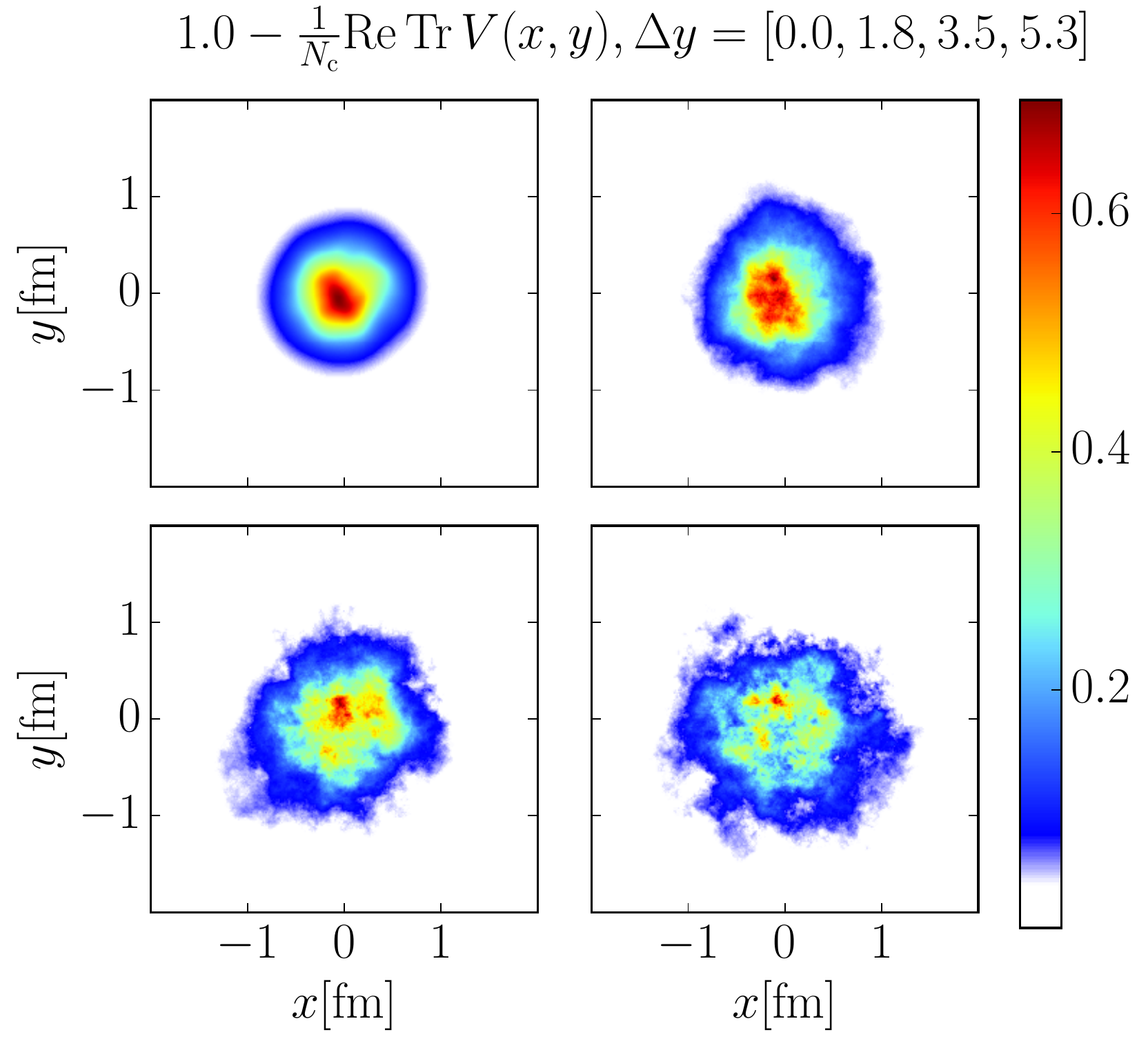} 
				\caption{Example of the proton density profile evolution over 5.3 units of rapidity with ultraviolet modes suppressed in the initial condition by $v=0.3\,\mathrm{GeV}^{-1}$.}
		\label{fig:tr_v_uv_damping_evolution}
		\end{minipage}
\end{figure*}

To access the proton size evolution, we study in more detail how the coherent spectrum evolves with $W$. The $|t|$ spectrum is sensitive to the Fourier transform of the density profile, as can be seen from Eq.~\eqref{eq:diff_amp}. Experimentally, the proton size is characterised by measuring the slope of the diffractive $|t|$ spectra, $B_G$, defined via
\begin{equation}
\frac{\der \sigma^{\gamma p \to J/\Psi p}}{\der t} \sim e^{-B_G |t|}.
\end{equation}
The diffractive slope has been measured as a function of the photon-proton center-of-mass energy $W$ by the H1 and ZEUS collaborations~\cite{Aktas:2005xu,Alexa:2013xxa,Chekanov:2002xi}.
This data clearly shows that the proton size grows as a function of $W$.  Recently, the ALICE collaboration has also measured $\gamma p$ scattering in ultraperipheral proton-nucleus collisions~\cite{TheALICE:2014dwa} and observed hints of the proton growth down to very small $x$. 

The proton size $B_G$ extracted from our calculation is shown in Fig.~\ref{fig:proton_size_diffraction} and compared with H1 and ZEUS data. Recall that we have fixed $B_p$ at the initial condition such that the resulting slope $B_G$ is approximately $3.8\gev^{-2}$ at $x=x_0$. The JIMWLK evolution results in a proton size evolution compatible with the HERA and ALICE data. For comparison, we also show the result obtained from the IPsat model, where the proton density profile does not change, and the only effect of the $W$ evolution is to increase the saturation scale $Q_s^2$. Because the profile function does not factorize in the IPsat model, there is a small residual $W$ dependence of $B_G$.

Even though both the fixed and running coupling evolution yield very similar descriptions of the charm structure function data, the proton size evolves faster when running coupling corrections are included. We explained this behavior in the discussion of Fig.\,\ref{fig:jpsi_t}: the running coupling suppresses short-wavelength modes relative to the longer-wavelength ones that are probed at small $|t|$. 

In these comparisons one should realize that both the computed and experimentally measured spectra are not exactly Gaussian (in $\Delta \approx \sqrt{-t}$), and thus it matters which range in  $|t|$ is fitted when the slope $B_G$ is extracted. The different experimental measurements have also different cuts, the ZEUS and older H1 datasets~\cite{Chekanov:2002xi,Aktas:2005xu} have only an upper cut $|t|<1.8\gev^2$ or $|t|<1.2\gev^2$, whereas the latest H1 analysis~\cite{Alexa:2013xxa} excludes the measured spectra below $|t|<0.1\gev^2$. In any case, there is little data available at small $|t|$ especially in the older analyses, thus we shall also impose a requirement $|t|>0.1\gev^2$ when extracting the slope $B_G$. 

Because the small $|t|$ region, important for $B_G$, is sensitive to large impact parameters, it is also especially sensitive to the infrared regulators, as discussed in Ref.~\cite{Mantysaari:2016jaz}. Thus, it is important to study the effect of $m$ on the extraction of $B_G$, which is done in Appendix \ref{appendix:mdep}.

To illustrate the growth of the proton and the change in its transverse structure, we show one example for the evolution of the trace of the Wilson lines, $1-\frac{1}{\nc}{\rm Re}\, {\rm tr}\, V(\xt)$ in Fig.\,\ref{fig:tr_v_ic}.  For comparison, the same evolution with the ultraviolet regulator $v=0.3\,{\rm GeV}^{-1}$ at the initial condition is shown in Fig.\,\ref{fig:tr_v_uv_damping_evolution}. The expected effect of the UV regulator of eliminating short range structures is clearly visible in the initial condition. JIMWLK evolution reintroduces short range structures and makes the proton grow as in the case of the MV model. 

\begin{figure}[tb]
\centering
		\includegraphics[width=0.5\textwidth]{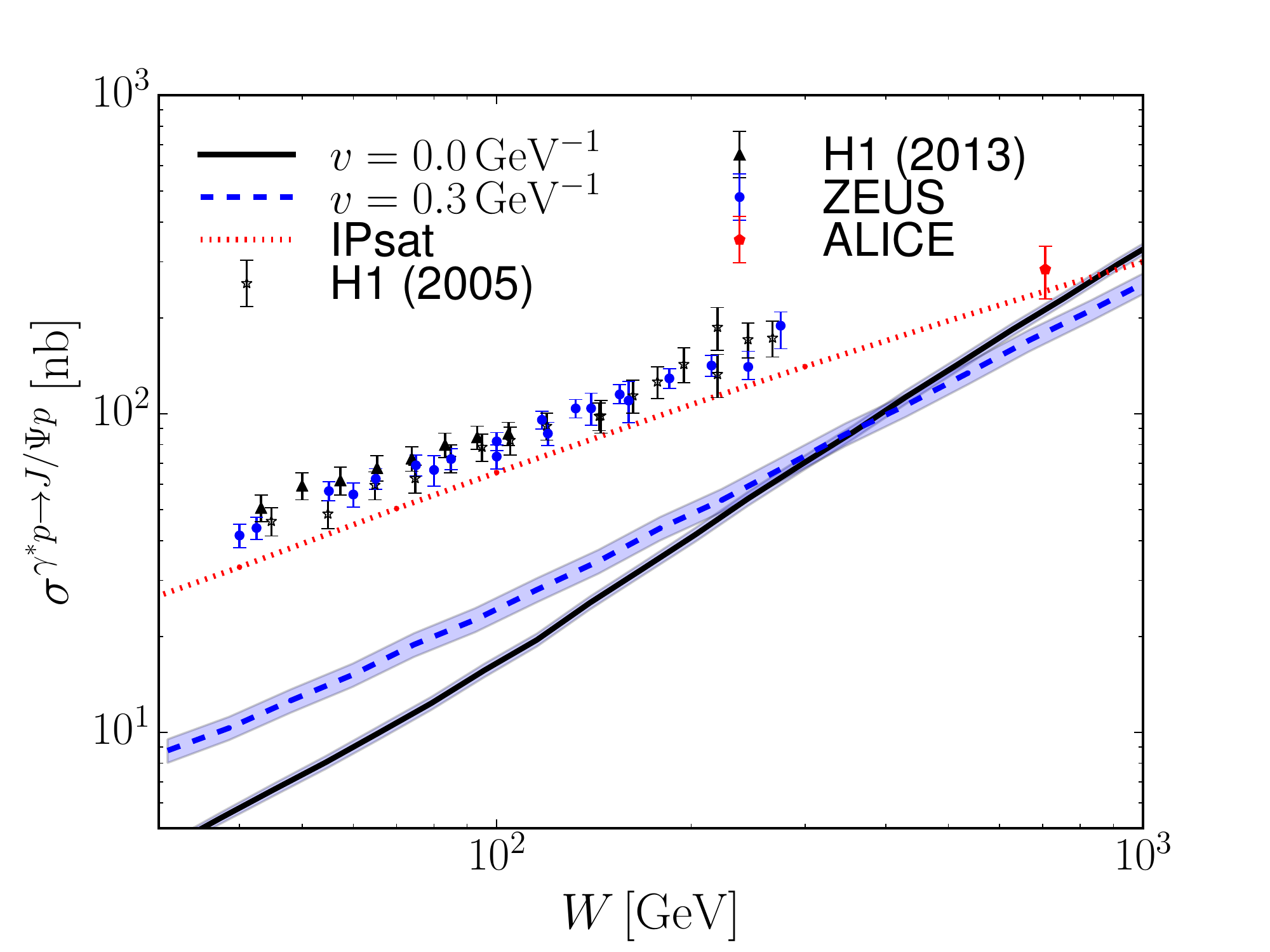} 
				\caption{Total coherent diffractive $J/\Psi$ photoproduction cross section as a function of center-of-mass energy $W$ and compared with H1 \cite{Aktas:2005xu,Alexa:2013xxa}, ZEUS \cite{Chekanov:2002xi} and ALICE \cite{TheALICE:2014dwa} data. }
		\label{fig:jpsi_totxs}
\end{figure}

Next, we investigate the effect of a finite UV regulator on the coherent diffractive $J/\Psi$ cross section. As shown in Fig.\,\ref{fig:jpsi_totxs} with the MV model initial condition ($v=0$), the calculated energy dependence of the total cross section is significantly faster than that of the experimental data. Including a UV regulator improves the energy dependence significantly, however, the result is still smaller than the data, for all $W$.  The reason is the lack of a soft non-perturbative contribution, just like it was for the total reduced cross section. We will return to the issue with the normalization in Sec.~\ref{sec:large_dipoles}.

The improvement of the energy dependence stems from the fact that in our fit to the charm reduced cross section, the case with UV regulator prefers a somewhat smaller $x$ evolution speed. In case of MV model initial condition, the fit tries to compensate more the too fast $Q^2$ evolution by having a smaller initial $g^4\mu^2$ and faster evolution speed.

\section{Fluctuating proton geometry}
\label{sec:fluctuating_geometry}

 \begin{figure}[tb]
\centering
		\includegraphics[width=0.5\textwidth]{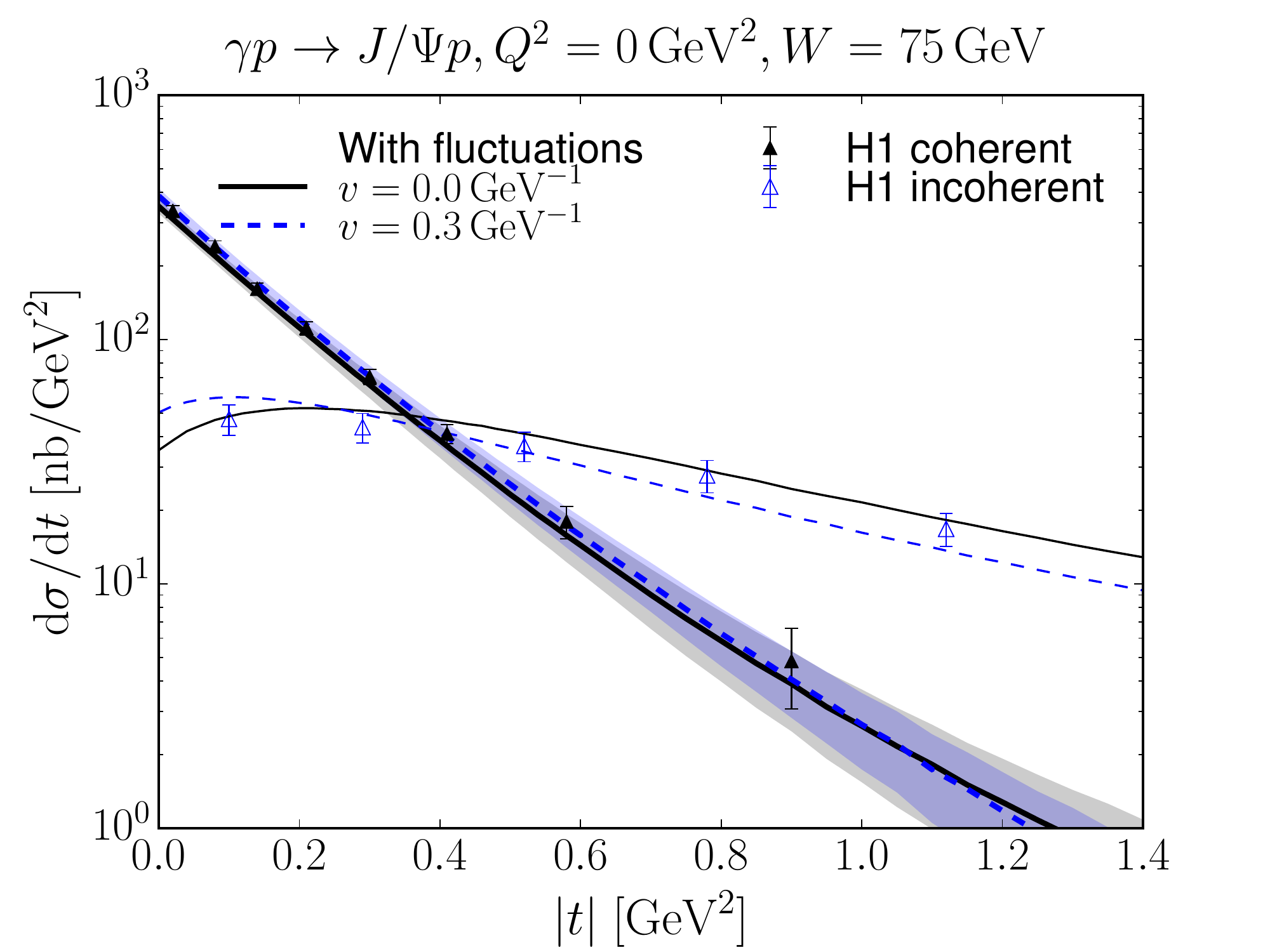} 
				\caption{Coherent (thick lines) and incoherent (thin lines) $J/\Psi$ photoproduction cross section at $W=75\,{\rm GeV}$ where the proton parametrization is fixed by the H1 data~\cite{Alexa:2013xxa}. Note that the proton color charge density is also fixed by the $J/\Psi$ data. 
				The results with and without UV damping in the initial condition are shown. 
				}
		\label{fig:fluctuating_spectra}
\end{figure}

 \begin{figure}[tb]
\centering
		\includegraphics[width=0.5\textwidth]{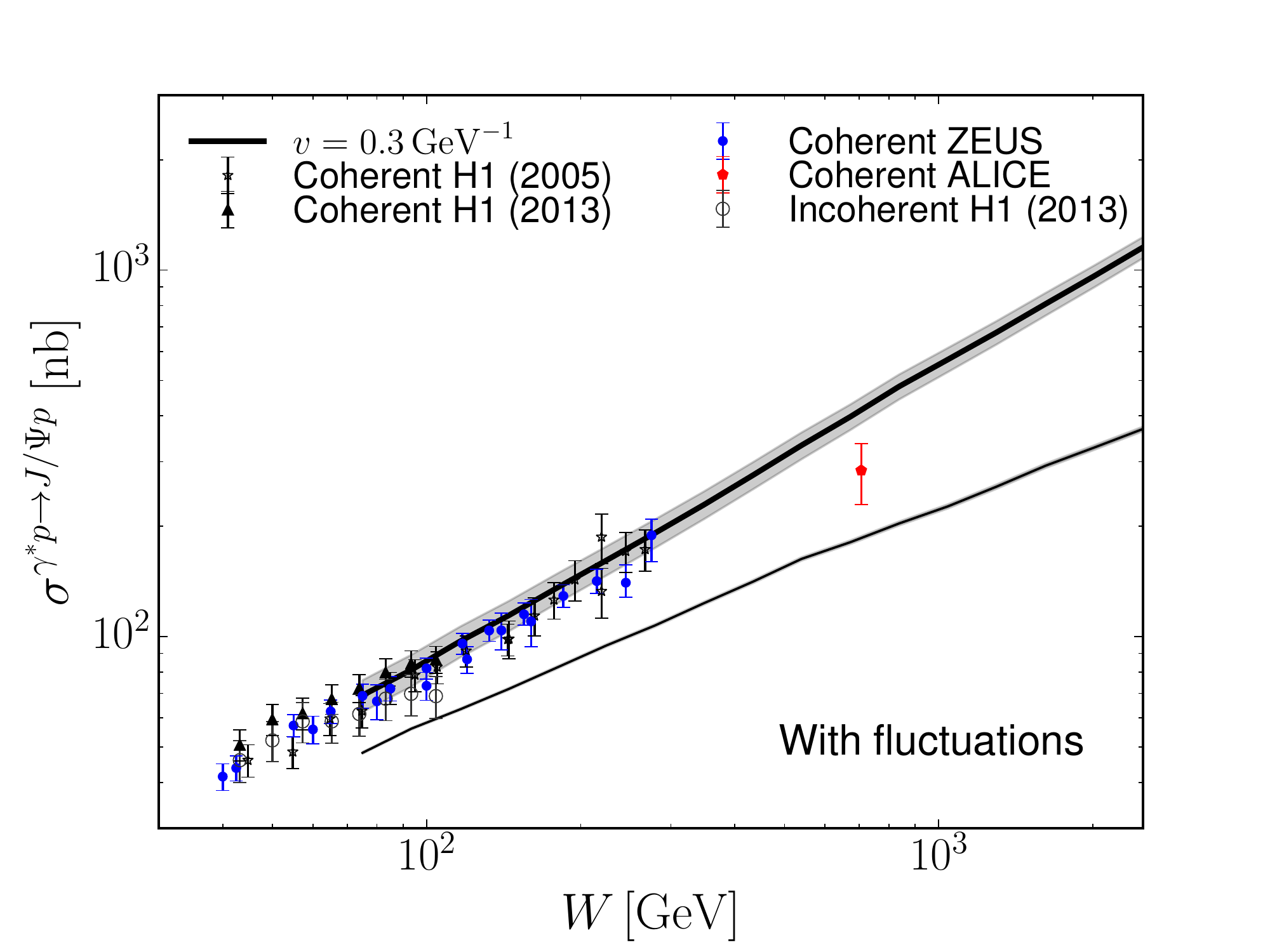} 
				\caption{Total coherent (thick lines) and incoherent (thin lines) $J/\Psi$ photoproduction cross section as a function of $W$. Here we only show results with ultraviolet damping ($v=0.3$) in the initial condition. Note that the proton color charge density normalization in the initial condition is fixed by the $J/\Psi$ data. We compare to experimental data from H1 \cite{Aktas:2005xu,Alexa:2013xxa}, ZEUS \cite{Chekanov:2002xi} and ALICE \cite{TheALICE:2014dwa}.
				}
		\label{fig:totxs_fluct}
\end{figure}
When we include fluctuations in the proton geometry and in the overall saturation scale, it becomes possible to study, in addition to the coherent cross section, also the incoherent process which measures the amount of event-by-event fluctuations. Here, similar to Ref.~\cite{Mantysaari:2016jaz}, we fix the parameters defining the fluctuating proton geometry ($B_{qc}$ and $B_q$) and the overall saturation scale controlled by $g^4\mu^2$ by requiring that we get a good description of the H1 spectra at $W=75\gev$~\cite{Alexa:2013xxa}. Because of the problem with non-perturbative contributions from large $r$ to the diffractive cross section, this normalization is different from what the fit to the reduced charm cross section would require. Consequently, the parameter set used in Refs.~\cite{Mantysaari:2016ykx,Mantysaari:2016jaz} also cannot reproduce the normalization of the charm production data.\footnote{Fortunately, some observables, like the incoherent to coherent cross section ratio, are rather insensitive to this normalization, as we will show below.}

We study the evolution of the diffractive cross sections towards large center-of-mass energies using JIMWLK evolution. We note that in this case the initial $x_0\approx 10^{-3}$ (as parameters are fixed by $J/\Psi$ data at $W=75\gev$, unlike previously when we studied the structure function data with the round protons). When solving the JIMWLK equation, we use the values for $\as$ and $m$ that were previously constrained by the reduced cross section data.

\begin{table}[tp]
\begin{center}
\begin{tabular}{c|c|c|c|c|c}
$v$ $[\gev^{-1}]$ & $g^4\mu^2$ $[\gev^2]$ & $B_{qc}$  $[\gev^{-2}]$ & $B_q$  $[\gev^{-2}]$ & $\sigma$ &  $\alpha_s$  \\
 \hline
0.0 & 2.8 & 3.2 &  0.5 & 0.5 & 0.21 \\
0.3 & 6.6 &  3.4 & 0.01 & 0.2 & 0.18 
 \end{tabular}
\end{center}
\caption{Parametrizations used to describe the HERA coherent and incoherent $J/\Psi$ production data at $W=75\gev$ corresponding to $x=0.0017$. The value for the strong coupling constant is obtained by fits to charm reduced cross section data as in Sec.~\ref{sec:structurefun}. Here the infrared regulator in the JIMWLK evolution is $m=0.2\gev$. Note that with ultraviolet damping $v=0.3\gev^{-1}$ the physical interpretation of the parameters is somewhat obscured.}
\label{table:fluct_parameters}
\end{table}%

The description of the H1 coherent and incoherent $J/\Psi$ photoproduction data is shown in Fig.~\ref{fig:fluctuating_spectra}. 
The parameters describing the proton geometry are constrained by requiring a good description of both coherent and incoherent $J/\Psi$ production data. Note that as our setup is slightly different than in our previous work~\cite{Mantysaari:2016jaz} (we are not using the IPsat model where the proton density function appears in the exponent of the dipole amplitude used to extract the saturation scale), we do not get exactly the same parameters even though we are comparing with the same dataset when using a similar initial condition with $v=0.0$. 

Similar to the case of the round proton, we also perform the analysis by using the initial condition with an ultraviolet regulator $v=0.3\,\mathrm{GeV}^{-1}$. 
As discussed earlier, this causes the short distance structure to be mostly filtered out and the proton effectively grows and becomes smoother. Consequently, when we use $v=0.3 \gev^{-1}$, we are forced to distribute color charges as very sharp peaks that are relatively far away from each other in transverse space. 
As the overall normalization $g^4\mu^2$ in this section is fixed by the $J/\Psi$ data at $W=75\gev$, consequently these parametrizations would overestimate the charm structure function (e.g. the parametrization with $v=0.3\gev^{-1}$ overestimates $\sigma_{r,c}$ by $40\%$), but the $x$ evolution speed is found to be compatible with the data.

The JIMWLK evolved $|t|$-integrated coherent and incoherent cross sections as functions of $W$ are shown in Fig.\,\ref{fig:totxs_fluct}, comparing to IPsat with a fluctuating proton and data for the diffractive cross sections from H1 \cite{Aktas:2005xu,Alexa:2013xxa}, ZEUS \cite{Chekanov:2002xi} and ALICE \cite{TheALICE:2014dwa}. While the growth of the coherent cross section with $W$ is stronger than the incoherent, the incoherent cross section never decreases with $W$ as was observed in the calculation in \cite{Cepila:2016uku}. Here the results at fixed coupling and with ultraviolet damping in the initial condition are shown, in which case the coupling constant obtained, $\as = 0.18$, constrained by the $x$-dependence of the DIS data, is approximatively compatible with the HERA measurements. The larger $\as=0.21$ which is a result of the fit with a pure MV model initial condition would result in too fast $W$ dependence of the total cross sections, as already seen in case of the round proton in Fig.~\ref{fig:jpsi_totxs}

\begin{figure*}[tb]
\centering
\begin{minipage}{0.48\textwidth}
	\centering
		\includegraphics[width=\textwidth]{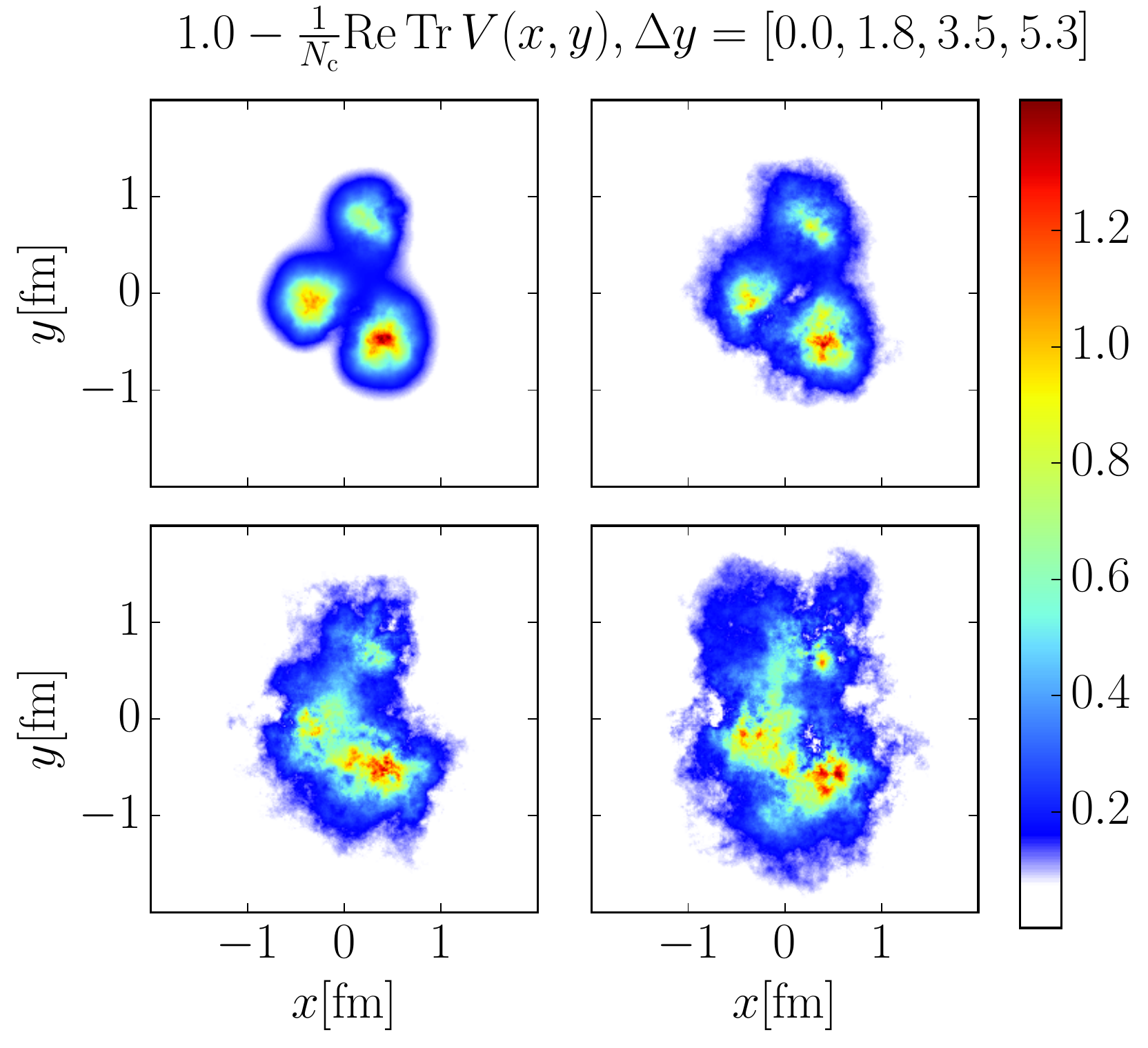} 
				\caption{An example evolution of the fluctuating proton shape over $5.3$ units of rapidity with no ultraviolet damping in the initial condition.	}
		\label{fig:evolution_fluct}
\end{minipage}
\quad
\begin{minipage}{0.48\textwidth}
\centering
		\includegraphics[width=\textwidth]{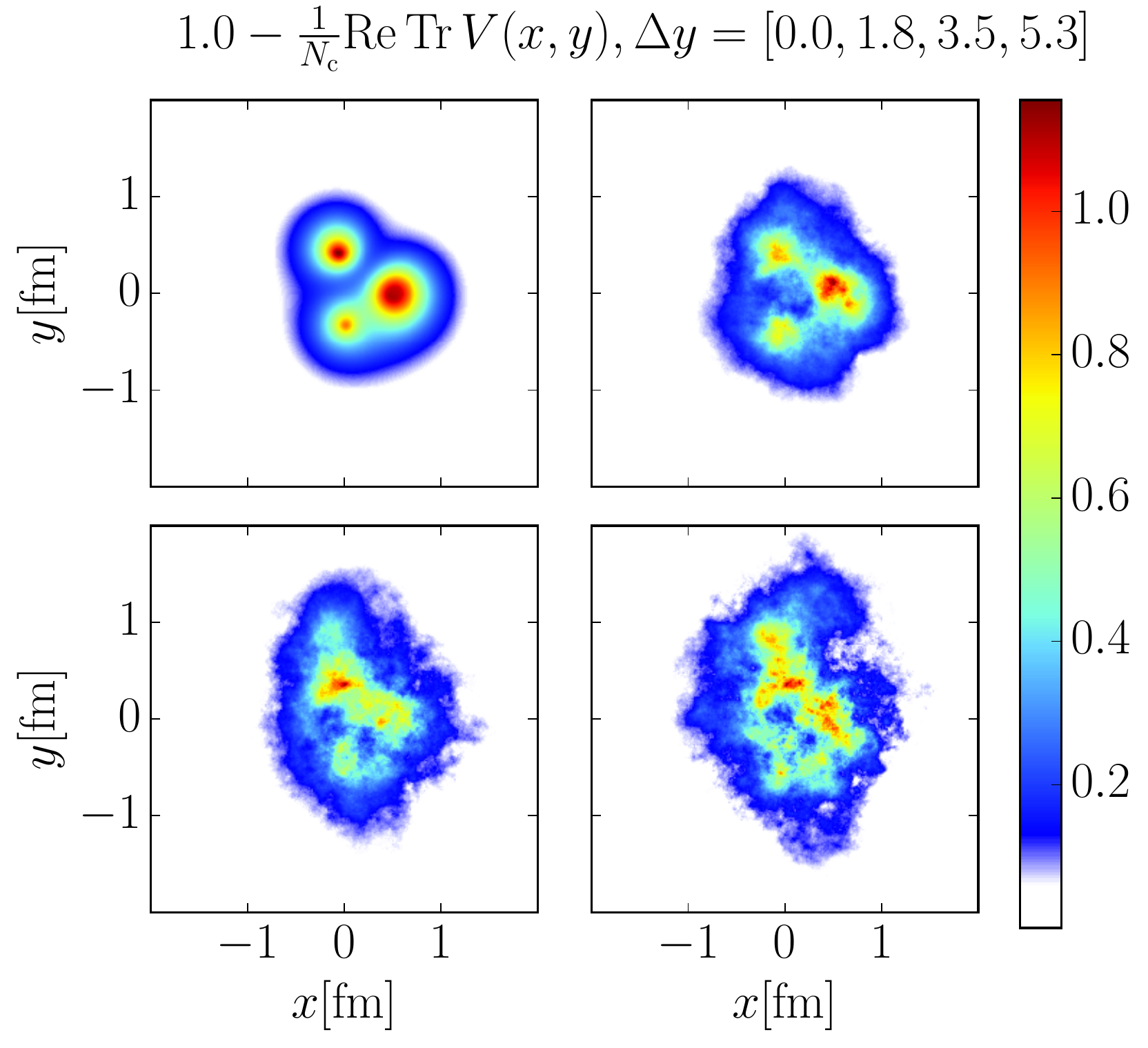} 
				\caption{An example evolution of the fluctuating proton shape over $5.3$ units of rapidity with ultraviolet damping $v=0.3\,\mathrm{GeV}^{-1}$ in the initial condition.	}
		\label{fig:evolution_fluct_uv}

\end{minipage}
\end{figure*}

In Figs.\,\ref{fig:evolution_fluct} and \ref{fig:evolution_fluct_uv} we show  one example for the rapidity evolution of the proton shape when starting from a fluctuating proton with three hot spots. The difference between the figures is that in Fig.~\ref{fig:evolution_fluct_uv} the ultraviolet damping factor is included in the initial condition, which removes short distance fluctuations. Comparing to Figs. \ref{fig:tr_v_ic} and \ref{fig:tr_v_uv_damping_evolution}, one notices that at large rapidities the shapes for initially round and fluctuating protons become similar. 

The behavior observed in Fig.\,\ref{fig:totxs_fluct} that the coherent cross section dominates at high energies, is expected. The reason is that if we start from proton configurations with large event-by-event fluctuations, the evolution is fastest in the regions where the local saturation scale is small, compared to the centers of the hot spot that start to reach the saturated region. This causes the hot spots to grow, and the proton effectively gets smoother, which reduces the fluctuations. Also, with increasing $Q_s$, the typical length scale of fluctuations becomes smaller, producing more domains, effectively decreasing the total amount of geometry fluctuations. Eventually, when the black disk limit is reached, the coherent cross section gets contributions from the whole transverse area of the proton where the dipole amplitude is saturated to unity. The incoherent cross section, on the other hand, can not receive any contribution in this region, and becomes only sensitive to the edge of the proton. 

\begin{figure}[tb]
\centering
		\includegraphics[width=0.5\textwidth]{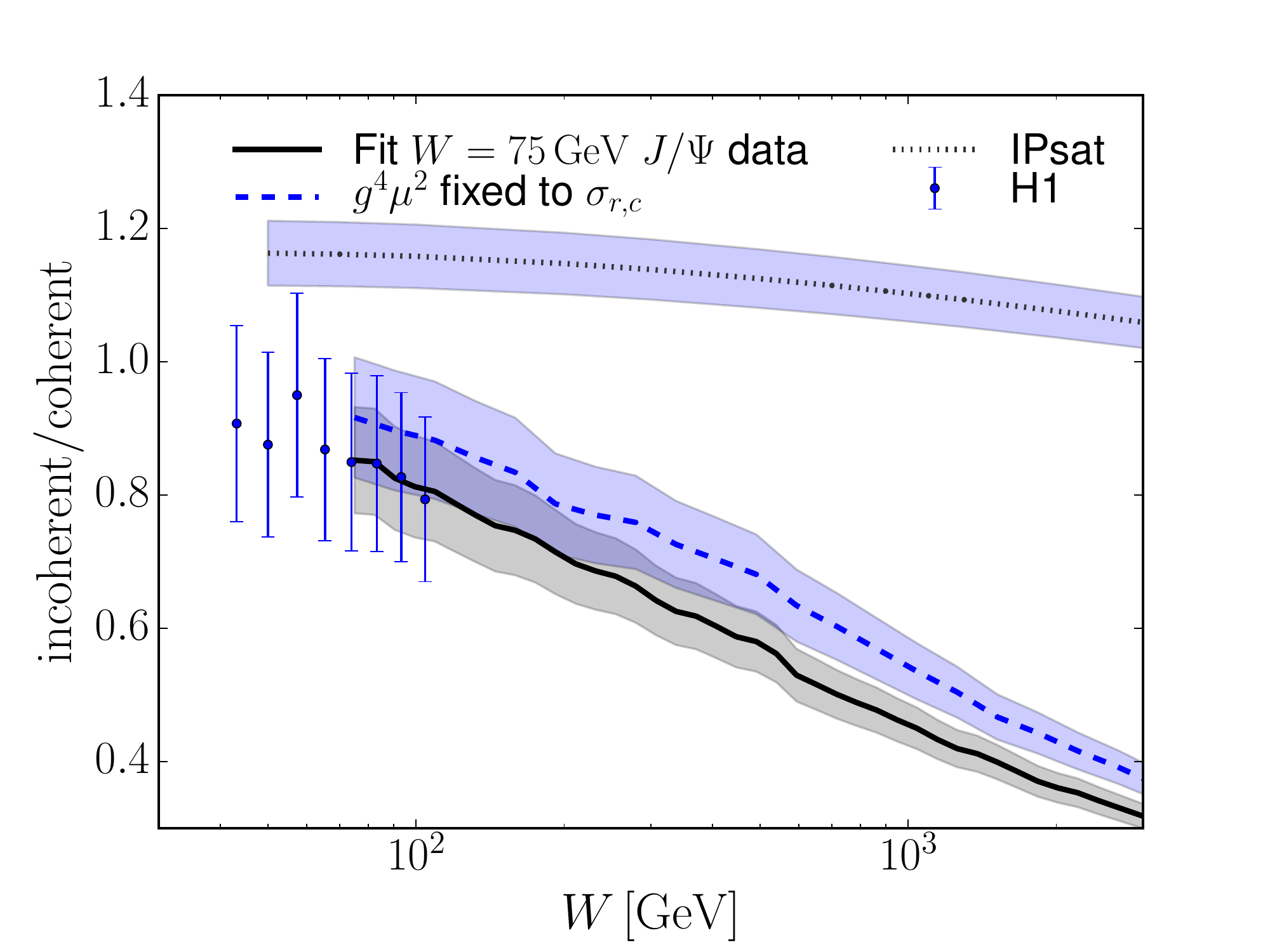} 
				\caption{Dependence of the incoherent-to-coherent cross section ratio on the color charge density normalization in the initial condition (for the solid line the $g^4\mu^2$ is fixed to $J/\Psi$ spectra, and for the dashed line it is fixed by the charm production cross section data). The results are compared with the HERA data~\cite{Alexa:2013xxa}. No ultraviolet damping is included here. The experimental uncertainties are computed assuming completely independent uncertainties for the coherent and incoherent cross sections.  }
		\label{fig:incohcohratio}
\end{figure}

This behavior is most clearly visible in the ratio of the incoherent to the coherent diffractive cross sections, shown in Fig.~\ref{fig:incohcohratio} and compared with the H1 data~\cite{Alexa:2013xxa}.  We present results for the choice of parameters that produce a good fit to the H1 spectra at $W=75\gev$ and compare to the  parameter set where $g^4\mu^2$ is adjusted to fit the charm reduced cross section. The ratio is shifted slightly when changing the parameters, but the $W$ dependence, which is a prediction based on JIMWLK evolution, is unaffected. 

For comparison, we will show results obtained using an IPsat model with fluctuating hot spot structure parametrized in Ref.~\cite{Mantysaari:2016jaz}. The proton structure parameters in that case are $B_{qc}=3.3\gev^{-2}$, $B_q=0.7\gev^{-2}$ and $\sigma=0.5$. 
The resulting cross section ratio 
 is much flatter as a function of $W$, because it lacks important physics, including the proton growth and evolution of the fluctuating sub-structure, and only the overall saturation scale depends on energy.  Additionally, we note that the cross section ratio is slightly above the data also at $W=75\gev$ where the parameters are constrained in Ref.~\cite{Mantysaari:2016jaz}. This is due to the steeper slope of the experimental coherent $t$ spectra than what can be reproduced by a proton with root mean square size $B_q + B_{qc}=4\gev^{-2}$ as used in Ref.~\cite{Mantysaari:2016jaz}, which describes data for $W=90\,{\rm GeV}$ well.

\begin{figure}[tb]
\centering
		\includegraphics[width=0.5\textwidth]{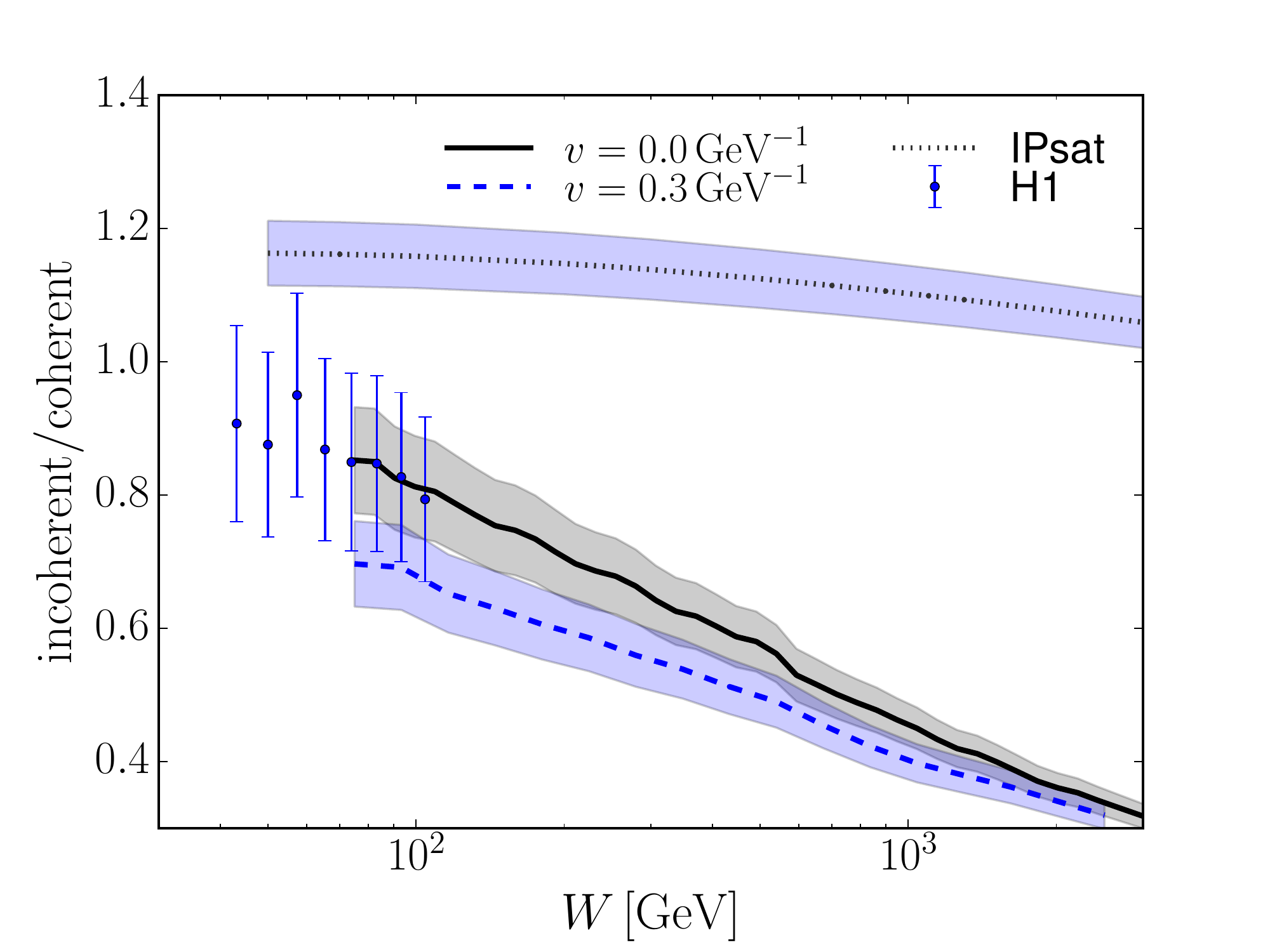} 
				\caption{Ratio of incoherent and coherent $J/\Psi$ photoproduction cross section as a function of center-of-mass energy $W$. In the dashed line the ultraviolet damping is included in the initial condition, and in both cases the parametrizations are fixed at the $W=75 \gev$ data.
				 The results are compared with the HERA data~\cite{Alexa:2013xxa}. }		\label{fig:incohcohratio_uv}
\end{figure}

In Fig.\,\ref{fig:incohcohratio_uv}, we compare to the case with UV damping in the initial state. Unsurprisingly, because the UV filter removes some fluctuations while keeping the overall size the same, the ratio of incoherent to coherent cross section is reduced. The evolution of the ratio with energy is similar, possibly slightly slower when UV damping is used. This makes the results with and without UV damping become more similar with evolution (this can also be observed in Figs.\,\ref{fig:amplitude_uv} and \ref{fig:amplitude_uv_q1_0}), which can be understood by the structure becoming dominated by JIMWLK effects and losing memory of the initial state.

\section{Large dipoles}
\label{sec:large_dipoles}
As shown in the previous sections, our results for inclusive structure functions and diffractive cross sections underestimate the experimental data, unlike computations relying on the IPsat parametrization. On the other hand, both approaches give a good description of the charm structure function (but we also note that the IPsat model can not describe the observed proton growth in $1/x$). This difference was explained by noticing that the behavior of the dipole amplitude for large (compared to the size of the proton and to the inverse of the saturation scale) dipoles is very different in the two frameworks.

As previously mentioned, in our model and at zero impact parameter, the quarks do not probe the densest part of the proton, but that at distances $r/2$ from the center. This is in contrast to IPsat where the saturation scale is probed at the impact parameter (the point between the quark and anti-quark). This difference becomes important when the dipole size is of the order of the proton size. In principle, this is the region where confinement scale physics should dominate, but no such phenomena are included in our model. In the IPsat model, large enough dipoles scatter with probability one, which can be seen as an effective description of confinement scale physics. 

In order to quantify how sensitive our results for different observables are to the description of large dipoles, and to estimate the importance of the confinement scale physics when describing the HERA data, we now study the dependence of the cross sections on an upper limit $r_\text{max}$, we impose on the dipole sizes to be included in the calculation. For simplicity, we do not include the geometry fluctuations in this analysis, as the results presented here only depend on the average shape of the proton. The modified initial condition with an ultraviolet damping factor $v=0.3\gev^{-1}$ constrained in Sec.~\ref{sec:structurefun} is used in this section. The results with an MV model initial condition are qualitatively similar. For a similar analysis in the IPsat model, see Ref~\cite{Mantysaari:2018nng}.

 \begin{figure}[tb]
\centering
		\includegraphics[width=0.5\textwidth]{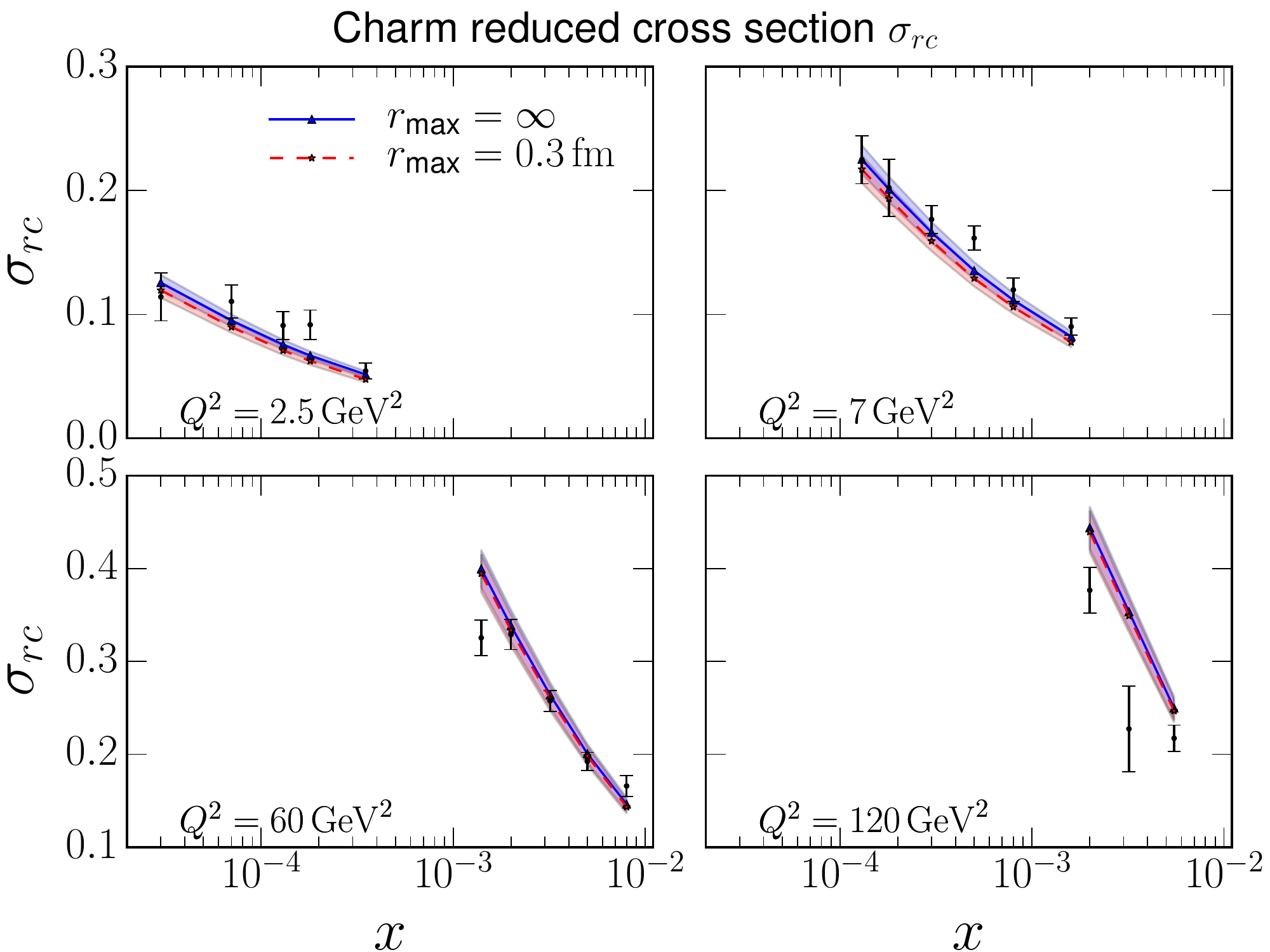} 
				\caption{Charm structure function computed with hard cutoff of $0.3\fm$ for the dipole size, compared to our original calculation.}
		\label{fig:sigmar_charm_rmax}
\end{figure}

The already encountered insensitivity of the charm structure functions to large dipoles is shown in Fig.~\ref{fig:sigmar_charm_rmax}. Here, the reduced charm cross section $\sigma_{rc}$ is compared with the result obtained by neglecting dipoles larger than $r_\text{max}=0.3\fm$. The resulting reduced cross section is found to be identical with and without this cutoff. This supports our approach to extract the free parameters (except for the proton size) by fitting the charm cross section data, as it is not sensitive to the physics at the confinement scale.

 \begin{figure}[tb]
\centering
		\includegraphics[width=0.5\textwidth]{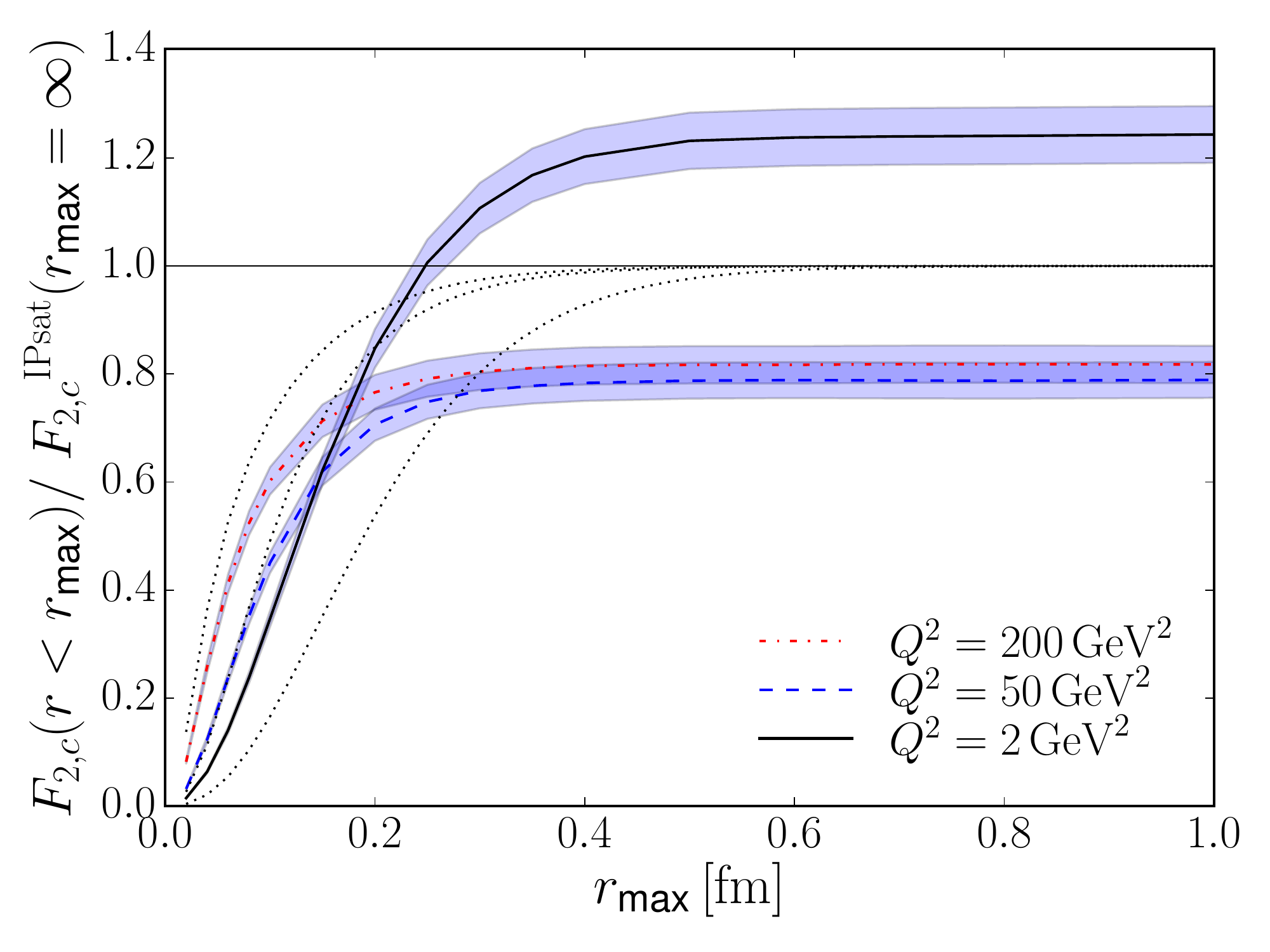} 
				\caption{Charm contribution to $F_2$ as a function of maximum dipole size at $x=0.01$ compared with the IPsat model result without $r_\text{max}$ cutoff. The dotted lines show the corresponding result obtained using the IPsat parametrization. }
		\label{fig:f2_charm_rmax}
\end{figure}

This is further demonstrated in Fig.~\ref{fig:f2_charm_rmax}, where the charm contribution to $F_2$ at $x=x_0=0.01$ is calculated at different values for the photon virtuality $Q^2$ as a function of $r_\text{max}$. More precisely, what is shown in Fig.~\ref{fig:f2_charm_rmax} is the amount of total $F_{2,c}$ recovered by including dipoles up to $r_\text{max}$, divided by the IPsat model result for $r_\text{max}=\infty$. The latter serves as a stand-in for the experimental value, as IPsat provides a very good description for this and the following observables.

For comparison, the results from the IPsat model\footnote{See also Ref.~\cite{Mantysaari:2018nng} for a more detailed analysis of the large dipole contributions in the IPsat parametrization.} with the same cutoff $r_{\rm max}$ (where large dipoles are given more weight) are shown as dotted lines. We find that even at  small $Q^2$, the total structure function is recovered already at $r_\text{max}=0.3\fm$.\footnote{Note that the full MV+JIMWLK result is slightly larger than the IPsat model result at $Q^2=2\,{\rm GeV}^2$ (where there is no experimental data available), leading to a ratio $>1$.} In the IPsat model, convergence happens at somewhat larger $r_{\rm max}\approx 0.4\,{\rm fm}$, which is expected given the larger dipole amplitudes at large $r$.

\begin{figure}[tb]
\centering
		\includegraphics[width=0.5\textwidth]{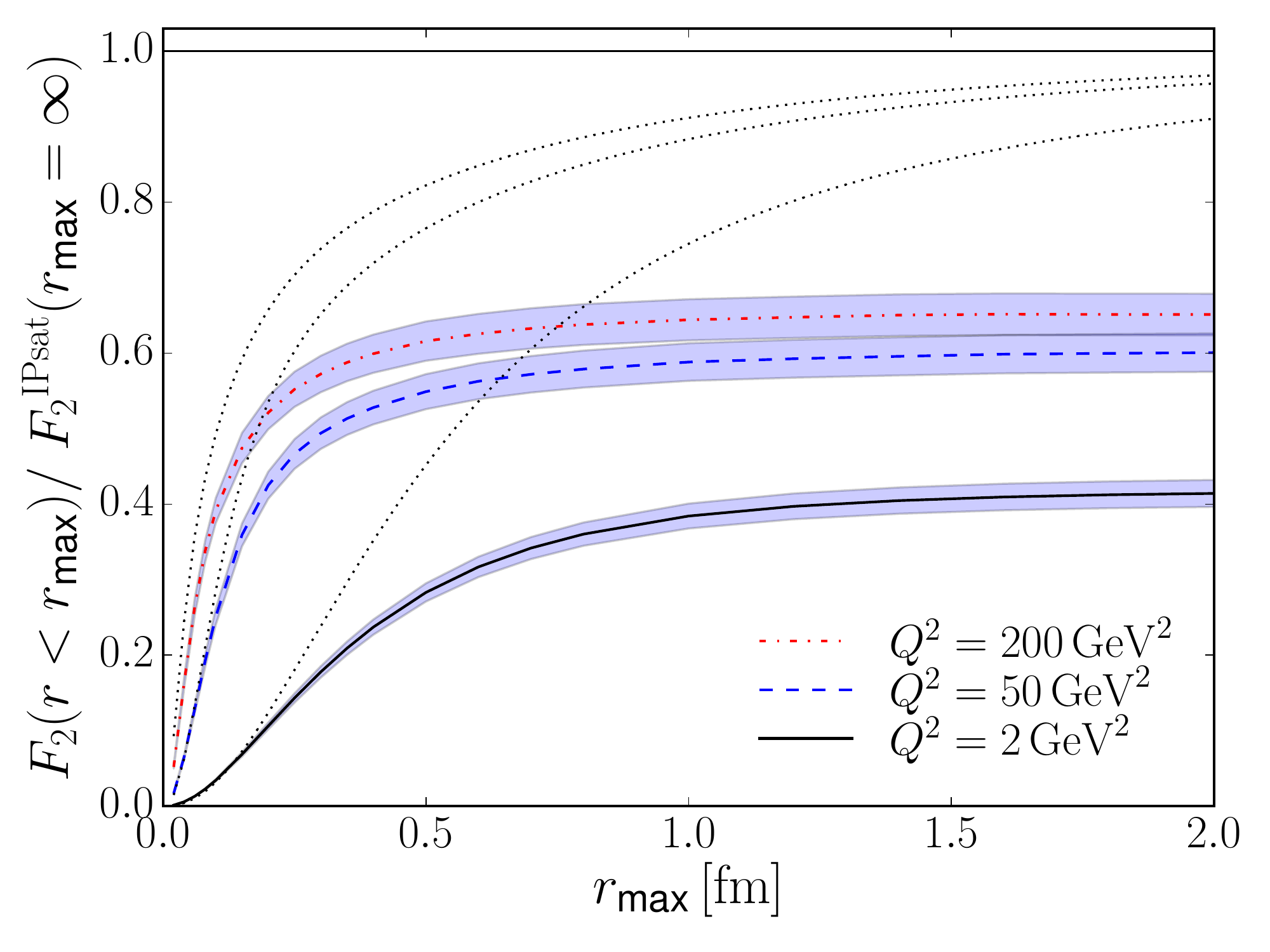} 
				\caption{$F_2$ as a function of maximum dipole size at $x=0.01$, compared with the IPsat model result without $r_\text{max}$ cutoff.   The dotted lines show the corresponding results obtained using the IPsat parametrization. }
		\label{fig:f2_rmax}
\end{figure}

A similar analysis for the total $F_2$ at $x=0.01$ is shown in Fig.~\ref{fig:f2_rmax}. Here the aligned jet contribution from large dipoles is especially large when we employ the IPsat parametrization, where even at large $Q^2=200\gev^2$ over $10\%$ of the total cross section originate from dipoles larger than $1\fm$. In our framework the contribution form dipoles larger than the inverse proton size are suppressed, thus the total $F_2$ is significantly smaller than in the IPsat model, and at large $Q^2$, the results converge at around $r_\text{max}\approx 0.3\,{\rm fm}$. However, at small $Q^2$ dipoles up to $1\fm$ contribute.

\begin{figure}[tb]
\centering
		\includegraphics[width=0.5\textwidth]{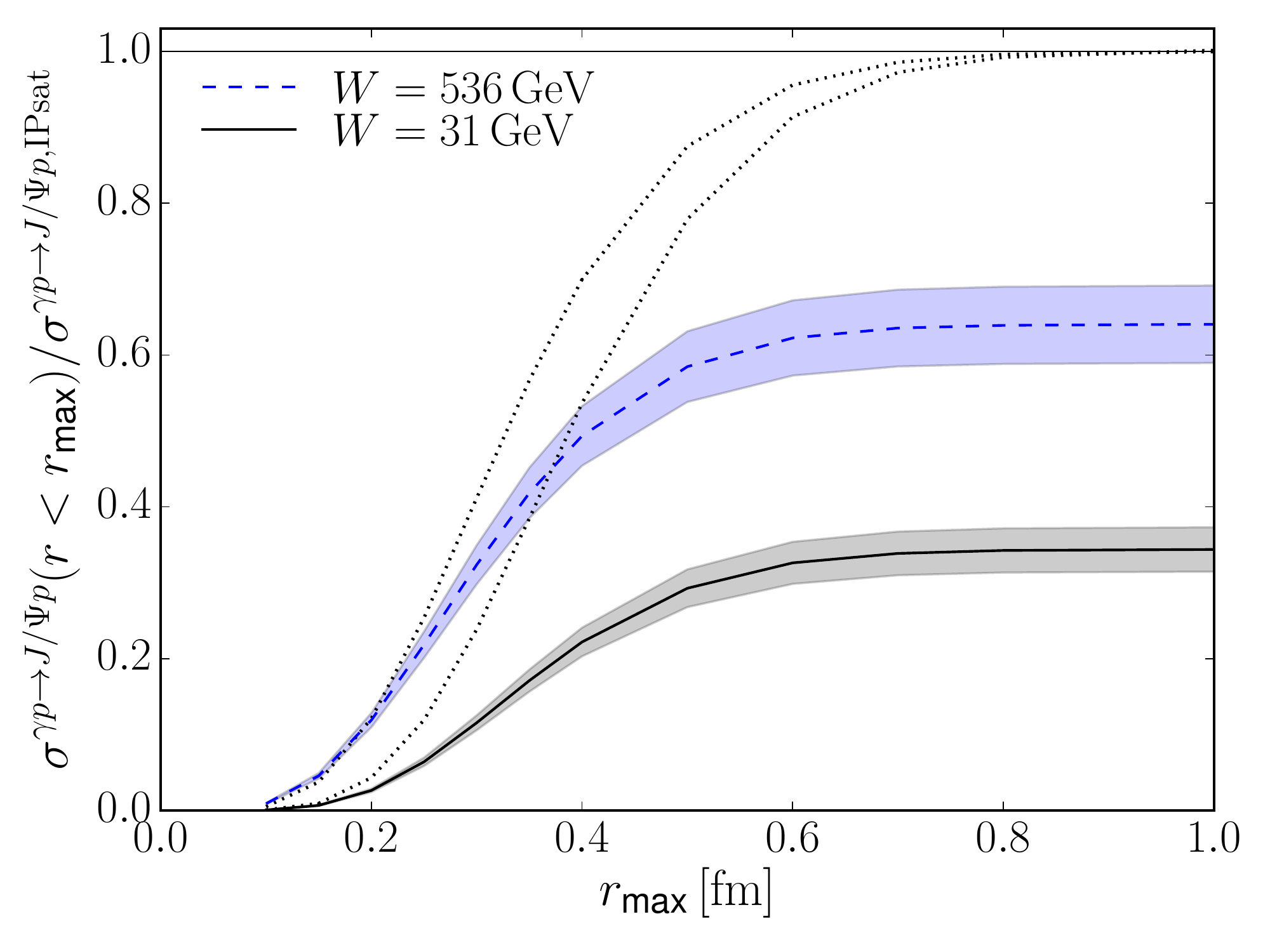} 
				\caption{Total $J/\Psi$ photoproduction cross section as a function of dipole size cutoff. The dotted lines show the corresponding results from the IPsat parametrization at the same $x$ (right at $\xpom=0.01$ and left at $\xpom=3.5\cdot 10^{-5}$).}
				
		\label{fig:diffractive_rmax}
\end{figure}

For the exclusive cross section, which depends on the square of the dipole amplitude, we expect an even stronger dependence on the cutoff $r_\text{max}$.
Additionally, the weight is given to different dipole sizes than in inclusive charm production even though in both cases only the charm dipoles contribute, as the dipole sizes are set by the overlap between the vector meson and the virutal photon wave functions, not the square of the photon wave function.
The total diffractive $J/\Psi$ photoproduction cross section as a function of $r_\text{max}$ is shown in Fig.~\ref{fig:diffractive_rmax}. 
For the smaller $W=31\,{\rm GeV}$ at the initial condition we find indeed a large difference of a factor of $2.9$ compared to the  IPsat result at $r_\text{max}=\infty$ (see also Fig.~\ref{fig:jpsi_totxs}), while at larger $W$ the difference is reduced.
The Bjorken-$x$ dependence of the $r_\text{max}$ dependence is also weak  and especially at small $W$ the normalization of the cross section is not described by our calculation. This is due to the missing non-perturbative contribution affecting dipoles larger than $\gtrsim 0.4\fm$, not included in our framework, but effectively present in the IPsat calculation.   
 We note that the vector meson wave function is rather uncertain and modifications could move weight away from large $r$, potentially improving the description within our framework.

\begin{figure}[tb]
\centering
		\includegraphics[width=0.5\textwidth]{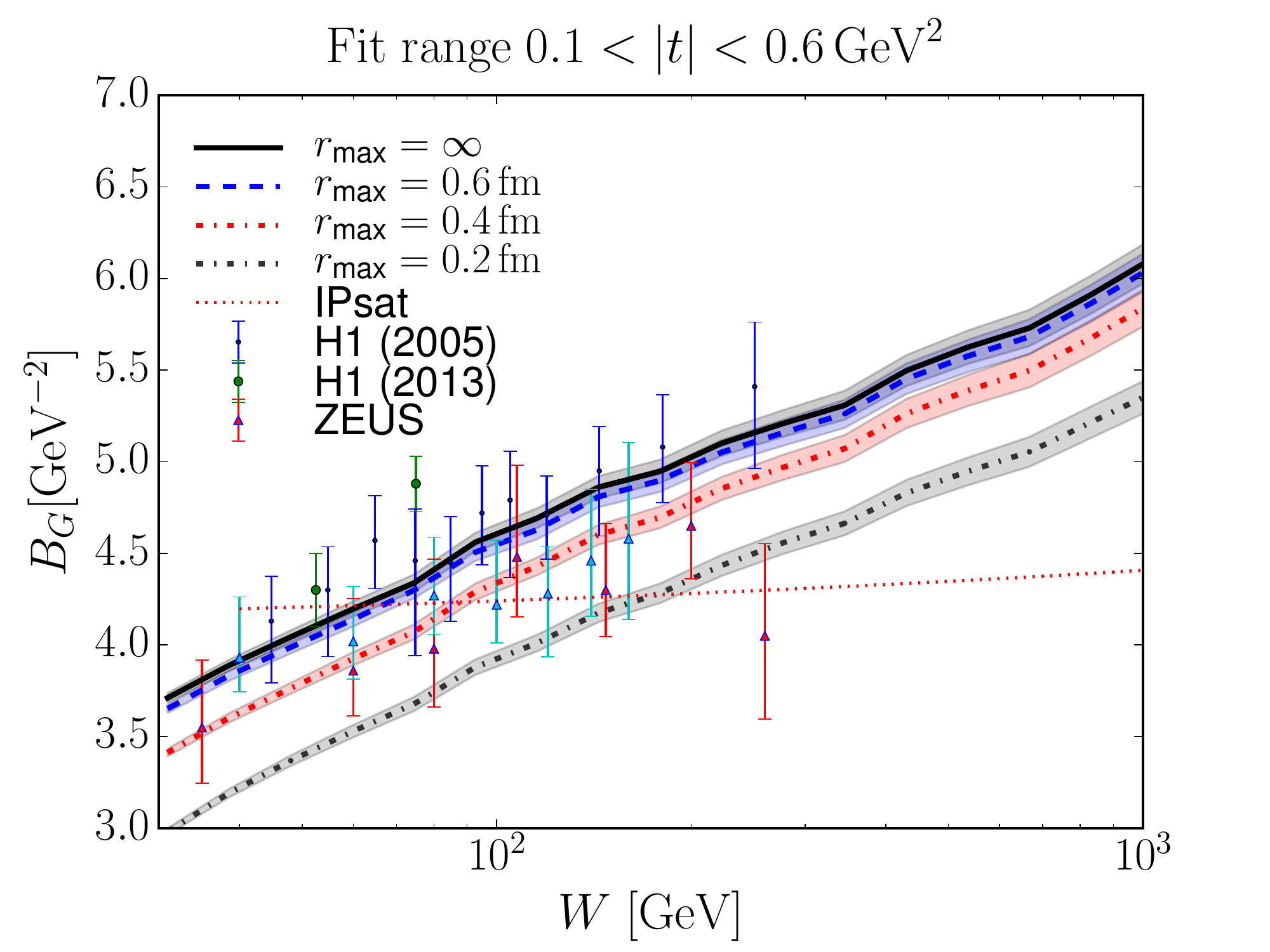} 
				\caption{Dependence on the dipole size cut $r_\text{max}$ of the diffractive slopes . 		}
		\label{fig:proton_size_diffraction_maxrdep}
\end{figure}

In  Fig.~\ref{fig:proton_size_diffraction_maxrdep} we show how the proton size extracted from exclusive $J/\Psi$ production depends on the cut on large dipoles. As long as the cut is not unreasonably small and $r_\text{max} \gtrsim 0.4\fm$, our results are compatible  with the HERA data. Especially the evolution speed with $W$ is independent of $r_{\rm max}$. However, this does not mean that the full non-perturbative result would have the same $W$ dependence, since the unknown soft contribution could lead to a modification.

 \begin{figure}[tb]
\centering
		\includegraphics[width=0.5\textwidth]{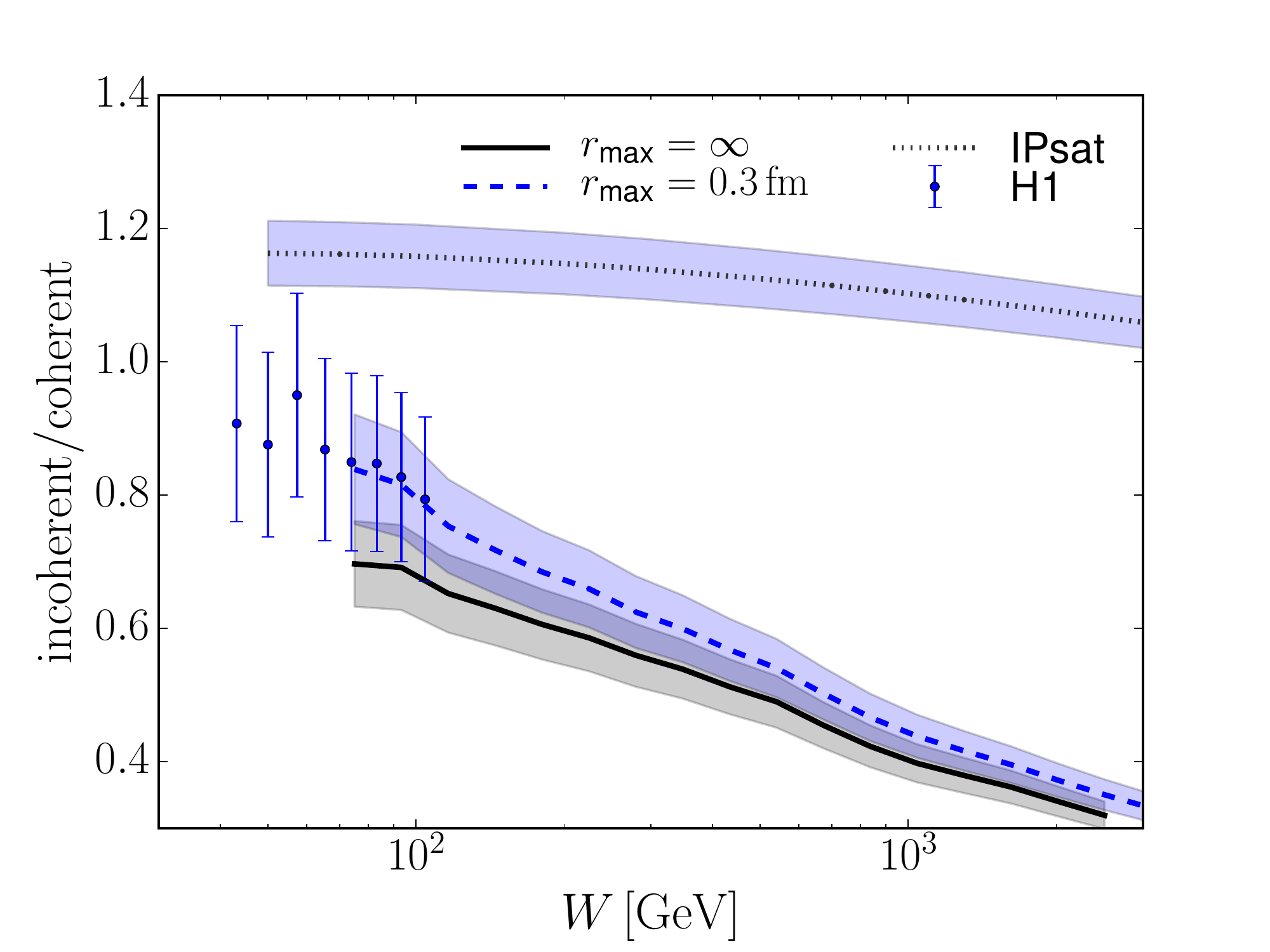} 
				\caption{Incoherent to coherent cross section ratio with and without cutoff for the large dipoles compared with the H1 data~\cite{Alexa:2013xxa}.		}
		\label{fig:incohcohratio_maxr}
\end{figure}

Finally in Fig.~\ref{fig:incohcohratio_maxr} we show how the incoherent to coherent cross section ratio depends on the contribution from large dipoles. The cross section ratio as a function of center-of-mass energy $W$ is shown using both the full solution to the JIMWLK evolution, and the result obtained by imposing a cut $r<r_\text{max}=0.3\fm$ for the dipole sizes. This cutoff changes the overall normalization of the cross sections, but the cross section ratio changes only slightly. In particular the energy dependence of the cross section ratio is independent of the large dipole cutoff.

The analysis done in this section has strong implications for the potential discovery of saturation effects in e+p collisions. We found that at small $r$ the perturbative description is valid and observables that exclude large dipoles can be well described. However, non-perturbative effects will set in when $r$ is on the order of $\Lambda_{\rm QCD}^{-1}$ and the differences between our model and saturation models like IPsat at large $r$ clearly demonstrate that this region is not under control. 
For realistically achieveable $x$ values at HERA energies, the dipole amplitude should be affected by this non-perturbative physics at values of $r$ that do not exceed $1/Q_s$ by much. Thus, the observables that are under control are not affected by saturation effects, and the observables that could be sensitive to saturation are not fully accessible in our  perturbative framework.

We thus conclude that access to saturation effects at currently realistic collider energies can only be achieved in collisions with heavy nuclei, where at a given energy, $Q_s^2$ is increased by $A^{1/3}$, with $A$ the mass number of the nucleus. Thus, saturation effects could be established at small enough $r$, where the perturbative treatment of the color glass condensate is still under good control.

\section{Conclusions and Outlook}
We have presented the first comparison of HERA data on structure functions and diffractive vector meson production to calculations involving the explicit numerical solution of the JIMWLK equations with impact parameter dependent MV model initial conditions.
We have used the charm reduced cross section and the slope of the $|t|$ spectrum for coherent $J/\Psi$ production to constrain the parameters of the model. Predicting the total reduced cross section as well as the coherent diffractive cross section (including its normalization) using these parameters leads to a significant underestimation of the experimental data. 
This discrepancy is rooted in the contribution from large dipoles, which is largely underestimated in our calculation. When the dipole misses the proton, the dipole amplitude goes to zero. This is not unreasonable, however, dipole sizes on the order of the proton size cannot be described perturbatively, because confinement effects become important. 
In fact, we find that limiting contributions from our perturbative calculations to $r<0.4\,{\rm fm}$ and adding a non-perturbative contribution for the soft part, which is represented by a simple toy model in this work, allows for a good description of the total reduced cross section.

For the IPsat model, good agreement is found with almost the entire range of HERA data~\cite{Rezaeian:2012ji}. This is because the large $r$ behavior of the dipole amplitude is qualitatively different from our result. In the IPsat model the dipole amplitude always approaches one at large $r$, leading to significant contributions to some observables from dipole sizes exceeding $1 \fm$. One could understand this contribution as an effective way to include non-perturbative physics, but it is not clear that the concept of a dipole amplitude at such large values of $r$ makes sense.

Observables that do not get contributions from very large dipole sizes, like the reduced charm cross section, can be well described within our framework. Other observables need to be supplemented with a non-perturbative contribution, which could be modeled via vector meson dominance or other more ad-hoc approaches. 

 We identified the ratio of incoherent to coherent diffractive $J/\Psi$ production cross sections and the energy dependence of $B_G$ as other observables that are robustly described in our framework, and are only weakly affected by contributions from large dipoles.
 
We have further analyzed the effect of running coupling and various infrared and ultraviolet regulators on the JIMWLK evolved observables. We find that the energy dependence of the proton size $B_G$, accessible via coherent diffractive vector meson production, is sensitive to running coupling effects as well as the infrared regulator, even when other parameters are retuned to reproduce the charm reduced cross section. This is because the proton size is more sensitive to infrared physics than other observables. The energy dependence of $B_G$ is little affected by contributions from large dipoles.

Generally, our results demonstrate that it is extremely difficult if not impossible to access saturation effects in e+p collisions at HERA energies, because either dipoles larger than $1/Q_s$ do not contribute or the perturbative framework breaks down for a given observable. We conclude that electron-ion collisions involving ions with large $A$ are needed to increase $Q_s$ at a given energy and study saturation in a theoretically controlled way. Consequently, an electron ion collider will be essential to access gluon saturation and study this interesting and complex regime of non-linear QCD in the laboratory.

\section*{Acknowledgments}
We thank A. Kurkela, T. Lappi, S. Schlichting, D. Takaki,  R. Venugopalan and P. Zurita for discussions. H. M. wishes to thank the Theoretical Physics Department at CERN and Nuclear Theory Group at BNL for hospitality during the completion of this work.
Computing resources of the National Energy Research Scientific Computing Center, which is supported by the Office of Science of the U.S. Department of Energy under Contract No. DE-AC02-05CH11231, and of the CSC -- IT Center for Science in Espoo, Finland, were used in this work.  This work was supported under DOE Contract No. DE-SC0012704. H. M. is supported by European Research Council, Grant ERC-2015-CoG-681707. BPS acknowledges a DOE Office of Science Early Career Award.

\appendix

\section{Dependence on the infrared cutoff in the JIMWLK kernel}
\label{appendix:mdep}

To study sensitivity on the infrared regulator $m$ in the JIWMLK kernel~\eqref{eq:jimwlk_m}, we study the description of the HERA data using different values for $m$. As the role of the infrared regulator is to suppress unphysical Coulomb tails, smaller values for the cutoff result in faster growth of the proton size, which is expected to affect the evolution speed of all observables. In this section, we use the MV model initial condition ($v=0$) and do not include geometry fluctuations for simplicity.

\begin{figure}[tb]
\centering
		\includegraphics[width=0.5\textwidth]{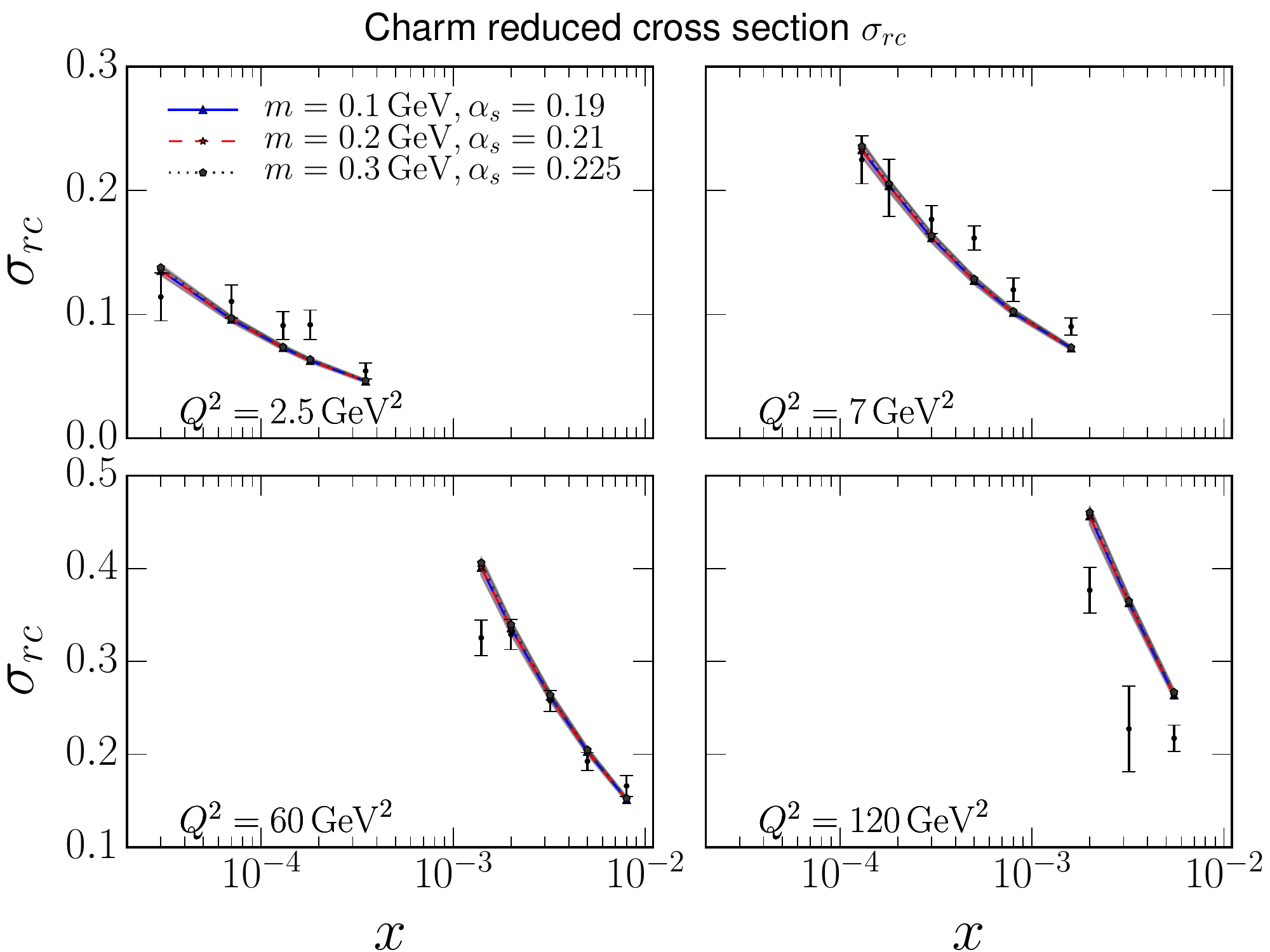} 
				\caption{Infrared cutoff dependence of the description of the charm structure function data. The fit quality is practically identical ($\chi^2/N \approx 4$).				}
		\label{fig:sigmar_charm_mdep}
\end{figure}

In practice we keep the initial color charge density $g^4\mu^2$ fixed and adjust the fixed coupling constant $\as$ separately for each infrared cutoff $m$ to get the best possible description of the charm production data. The results are shown in Fig.~\ref{fig:sigmar_charm_mdep}.

\begin{figure*}[tb]
\centering
\begin{minipage}{0.48\textwidth}
		\includegraphics[width=\textwidth]{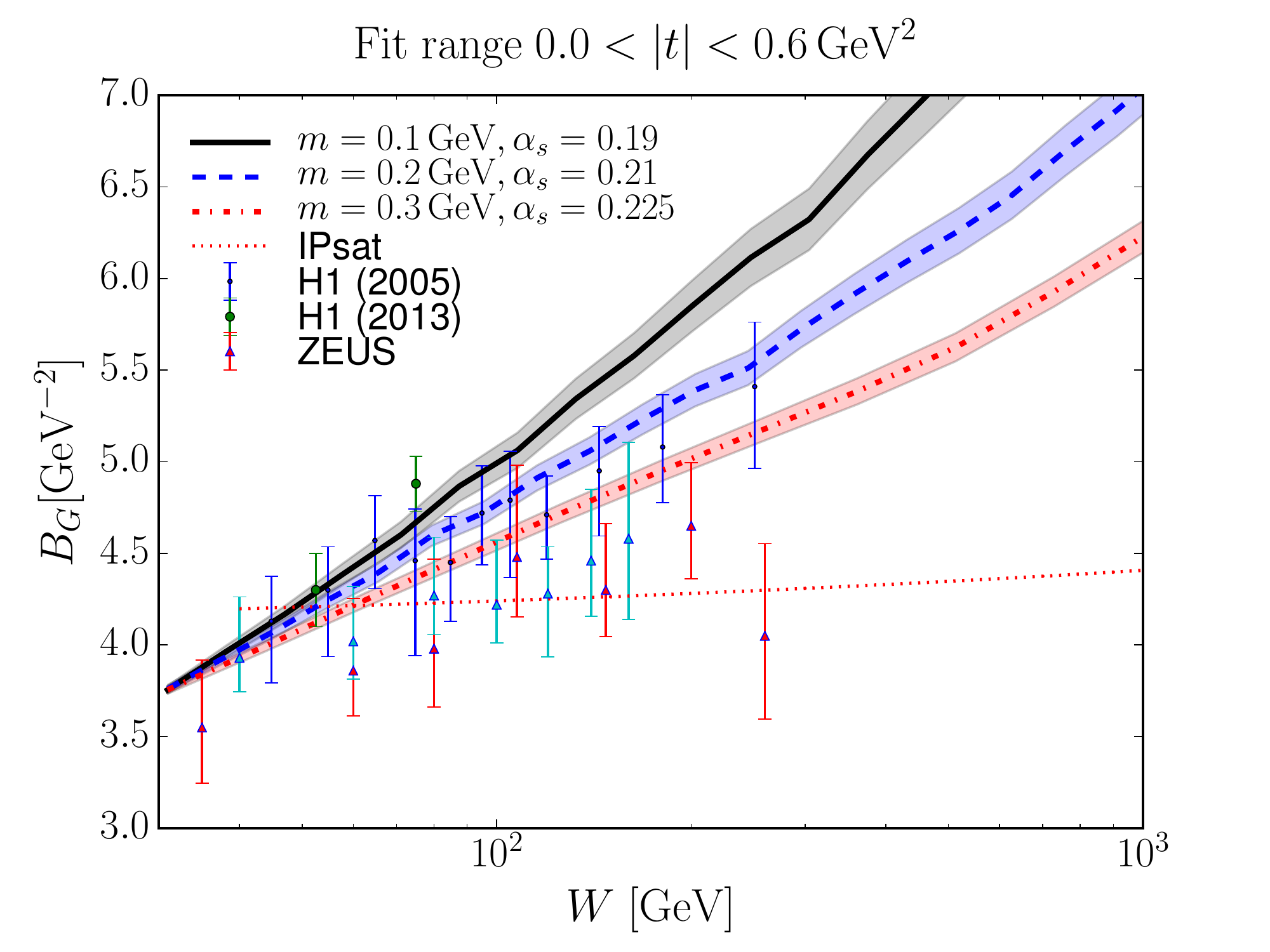} 
				\caption{Slope of the coherent diffractive cross section as a function of center of mass energy. The slope is extracted by fitting the computed spectra down to $|t|=0$.
				}
		\label{fig:slope_mdep_mint_0}
\end{minipage}
\quad
\begin{minipage}{0.48\textwidth}
\centering
		\includegraphics[width=\textwidth]{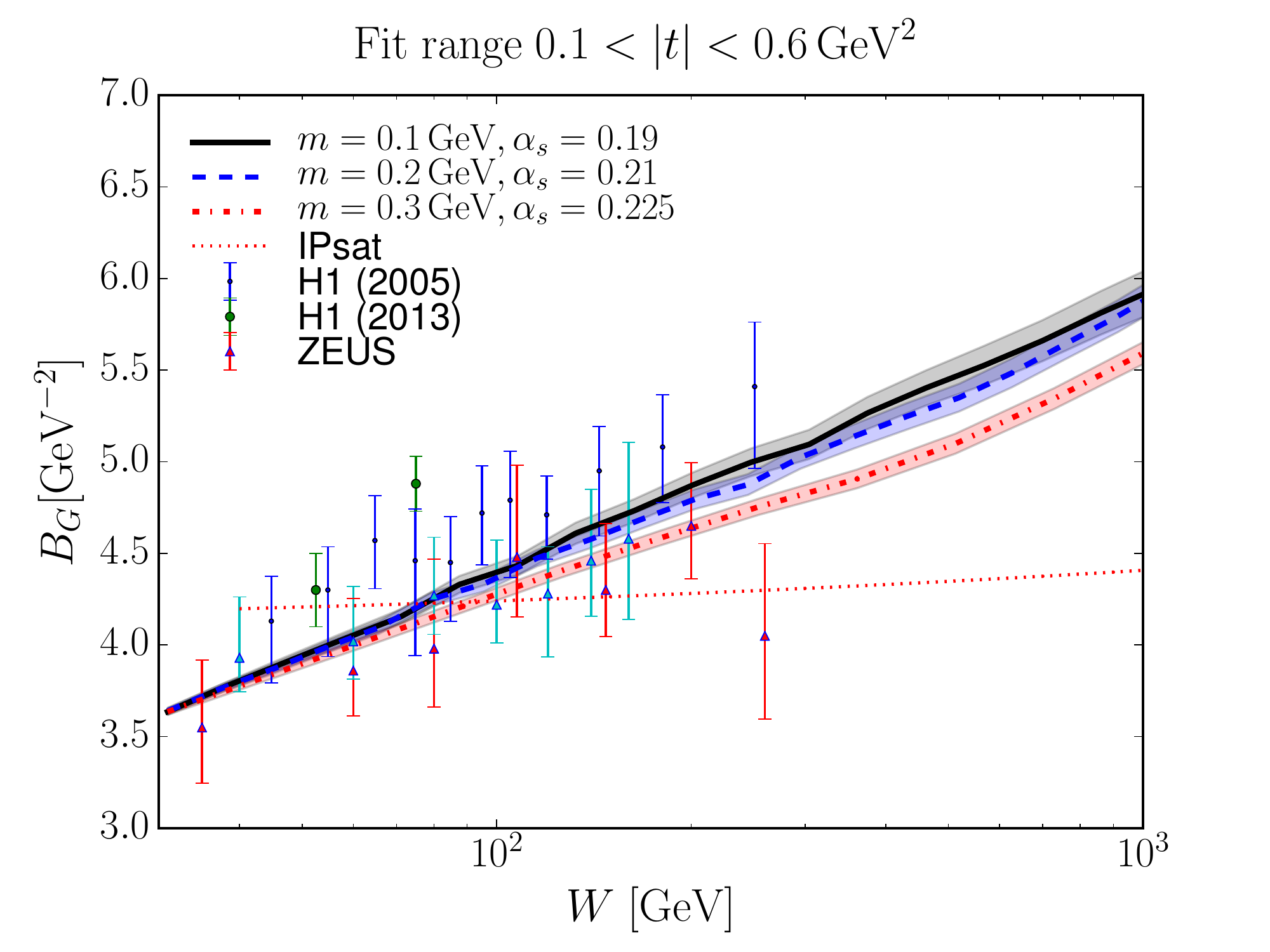} 
				\caption{Slope of the coherent diffractive cross section as a function of center of mass energy. The slope is extracted by fitting the computed spectra above $|t|=0.1\,\gev^2$.
				}
		\label{fig:slope_mdep_mint_0.1}
\end{minipage}
\end{figure*}

After retuning $\alpha_s$ to describe the charm reduced cross section, the effect of the infrared regulator on the proton size parameter $B_G$, measured in diffractive scattering, is shown in Figs.~\ref{fig:slope_mdep_mint_0} and \ref{fig:slope_mdep_mint_0.1}. We find that especially the small-$|t|$ part of the spectra is sensitive to the infrared regulator, leading to a strong $m$ dependence of the energy dependence of $B_G$ shown in Fig.\,\ref{fig:slope_mdep_mint_0}. However, if one follows the experimental cuts as closely as possible and excludes the region $|t| < 0.1~\gev^2$, the results are almost independent of $m$, as shown in Fig.\,\ref{fig:slope_mdep_mint_0.1}.

\begin{figure}[tb]
\centering
		\includegraphics[width=0.5\textwidth]{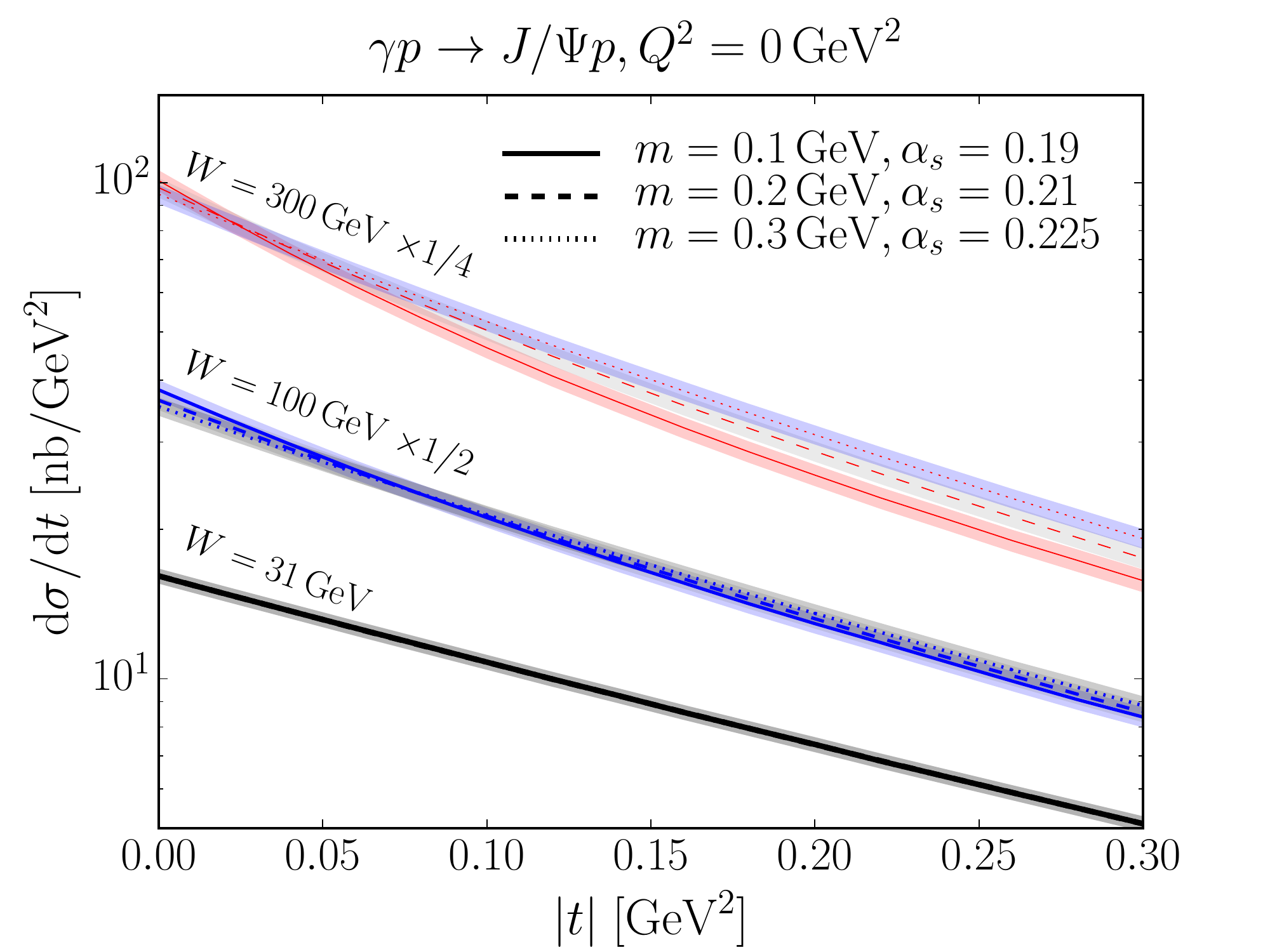} 
				\caption{$J/\Psi$ production at small $t$ and at fixed coupling computed with different infrared couplings $m$ in the JIMWLK evolution).}
				
		\label{fig:jpsi_smallt_mdep}
\end{figure}

To explain the $m$ dependence in the extracted diffractive slope, we show the coherent J/$\Psi$ production cross section as a function of $|t|$ in Fig.~\ref{fig:jpsi_smallt_mdep}. It is clear that if the infrared regulator is small, the long-distance modes cause the proton to grow much faster, creating an enhancement at small $|t|$. Also, in that case the spectrum is far from Gaussian, and extraction of the diffractive slope depends strongly on the $|t|$ range used in the analysis. We note that small enhancement at small $|t|$ (compared with purely Gaussian spectra extrapolated from larger $|t|$ to zero momentum transfer) can be seen in the latest HERA data at $W=75\gev$~\cite{Alexa:2013xxa}. 

\bibliographystyle{JHEP-2modlong.bst}
\bibliography{refs}

\end{document}